\documentclass[
 reprint,
superscriptaddress,
 amsmath,amssymb,
 aps,
prb,
]{revtex4-2}
\usepackage{graphicx}% Include figure files
\usepackage{dcolumn}% Align table columns on decimal point
\usepackage{bm}% bold math
\usepackage{hyperref}% add hypertext capabilities
\usepackage{comment}

\usepackage[normalem]{ulem}

\usepackage{xcolor}

\newcommand{\qv}{\mathbf{q}}
\newcommand{\av}[1]{\left\langle #1\right\rangle}

\begin{document}

%\preprint{APS/123-QED}

\title{Magnons in multiorbital Hubbard models, from Lieb to kagome}

\author{Teng-Fei Ying}
\affiliation{NanoLund and Division of Mathematical Physics, Department of Physics, Lund University, Lund, Sweden}
\affiliation{Thrust of Advanced Materials \& Quantum Science and Technology Center,
The Hong Kong University of Science and Technology (Guangzhou), 511453, Guangzhou, China}

\author{Hugo U. R. Strand}
\affiliation{School of Science and Technology, Örebro University, SE-701 82 Örebro, Sweden}

\author{Benjamin T. Zhou}
\affiliation{Thrust of Advanced Materials \& Quantum Science and Technology Center,
The Hong Kong University of Science and Technology (Guangzhou), 511453, Guangzhou, China}

\author{Erik G. C. P. van Loon}%
\affiliation{NanoLund and Division of Mathematical Physics, Department of Physics, Lund University, Lund, Sweden}
\affiliation{Wallenberg Initiative Materials Science for Sustainability, Department of Physics, Lund University, Lund, Sweden}

\date{\today}
 
\begin{abstract}
We investigate the magnetic orders and excitations in a half-filled Hubbard model that continuously interpolates between the Lieb and kagome lattices. 
Using self-consistent Hartree-Fock approximation combined with the Bethe-Salpeter equation in the random phase approximation, we map the $U-t'$ phase diagram of the Lieb-kagome lattices and compute real-frequency transverse spin susceptibilities, i.e., the magnon spectra. Dynamical mean-field theory and exact diagonalization are used as complementary checks of static magnetic phases.
In particular, besides gapless Goldstone magnons, the symmetry-broken ferrimagnetic and antiferromagnetic phases also exhibit gapped Higgs magnon bands, which originate from amplitude fluctuations in the order parameter characterizing spontaneous symmetry breaking.

\end{abstract}

\maketitle

\section{Introduction}

A hallmark of strongly correlated electron systems is the variety of phases that can appear as a result of electron-electron interactions. Out of the simplest single-orbital Hubbard model, exotic metallic, insulating, superconducting, and magnetic phases of matter can emerge~\cite{Qin22,Arovas22}. Going beyond a single-orbital Hubbard model, interactions lead to additional phases such as checkerboard charge-density waves~\cite{Hirsch84}, exciton condensates~\cite{Geffroy19}, and Hund's metals~\cite{Georges13}.

Ordered phases are generally characterized by a (static) order parameter, whereas changes in itinerancy are directly reflected in the single-particle excitation spectrum. On the other hand, the momentum-resolved dynamical susceptibility provides important complementary information, especially in the long-wavelength limit ($q\to0$). Notable examples include the disappearance of low-energy density fluctuations at a metal-insulator transition~\cite{Hafermann14,vanLoon14}, the emergence of gapless Goldstone modes in an exciton condensate~\cite{Geffroy19} and neutral inter-band collective excitations in multi-orbital systems, the so-called Pines' demon~\cite{husain2023pines}. 

\begin{figure}
    \centering
    \includegraphics[width=0.95\linewidth]{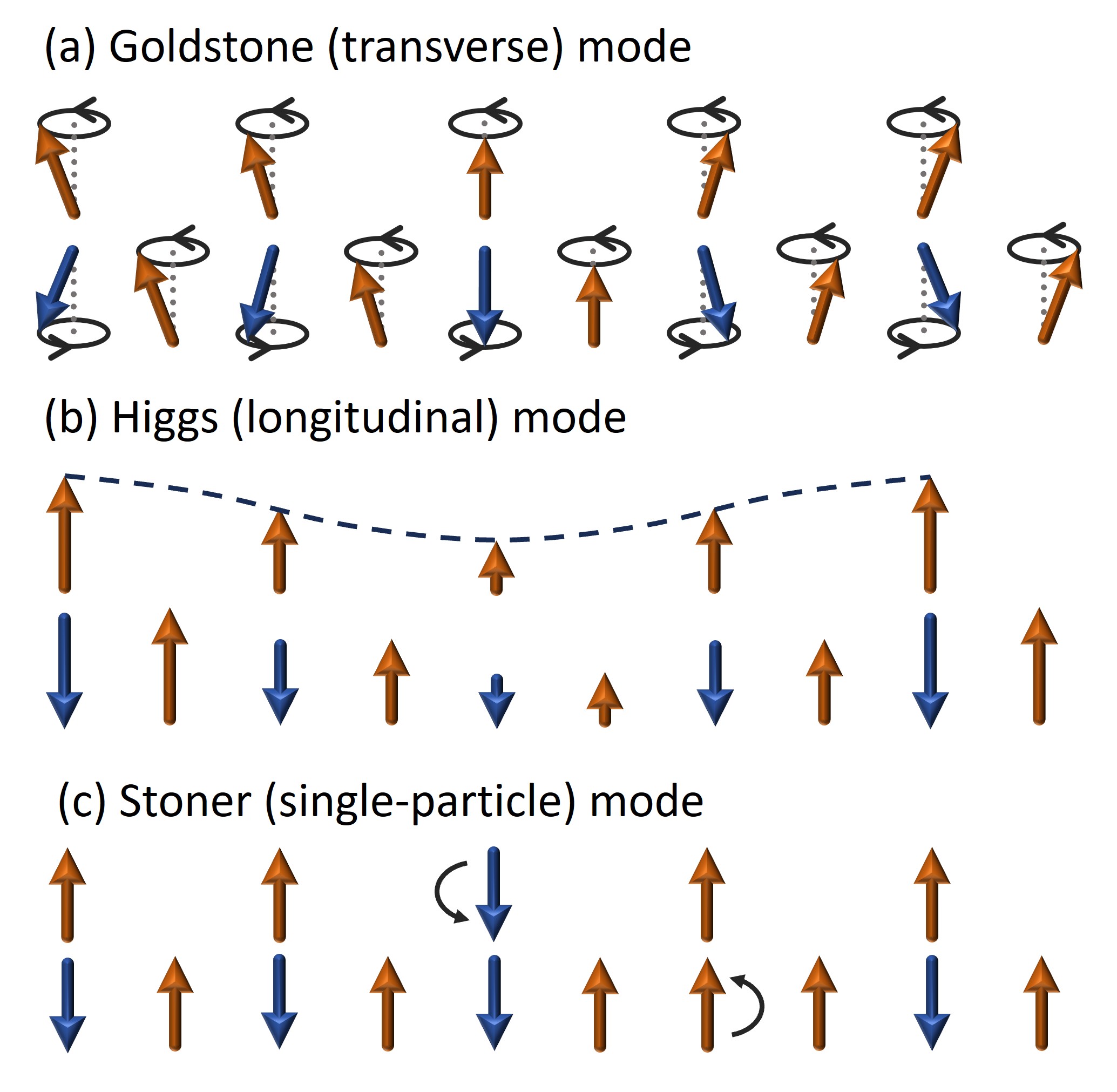}
    \caption{Illustrations of three types of magnetic excitations in the Lieb lattice: (a) the Goldstone mode, corresponding to a gapless magnon mode; (b) the Higgs mode, representing a gapped (amplitude) magnon from fluctuation of the order parameter; and (c) the Stoner mode (pair), describing single-particle spin-flip excitations. }
    \label{fig:illustration}
\end{figure}

For magnetic susceptibilities, which will be the main focus of this work, the type of magnetic order is essential for the structure of the dynamic susceptibility. In a magnetically ordered system, there are Goldstone modes (magnons) with vanishing energy at long wavelengths corresponding to the rotation of the order parameter. Their dispersion reflects the type of magnetic order: quadratic in ferromagnets, linear in antiferromagnets and altermagnets~\cite{Smejkal23}. For a paramagnet, the Goldstone mode is called a paramagnon and it also has linear dispersion. These Goldstone modes in electronic models are qualitatively similar to Heisenberg-like models~\cite{Eriksson17}, although effects such as doping dependence~\cite{Boehnke12,Musshoff21} cannot be directly captured in the Heisenberg model. In addition to the gapless Goldstone mode, there are two basic types of gapped magnetic excitations, as illustrated in Fig.~\ref{fig:illustration}. The Higgs (amplitude) mode~\cite{pekker2015amplitude} corresponds to the modulation of the order parameter and its energy is indicative of the free energy landscape of the order parameter, whereas the Stoner mode is an excitation of an electron from the majority to the minority band, with an energy given by the exchange splitting.  In multi-band models, the possibility of electronic excitations from occupied minority to empty majority bands makes it possible to have minority magnons~\cite{Skovhus24} and anti-Stoner modes~\cite{paischer2023nonlocal,paischer2024electronic}. Finally, altermagnets have chiral splitting~\cite{Smejkal23,MaierOkamoto23,sodequist24,Liu24}, in the sense that $\chi^{+-}(\omega,\qv) \neq \chi^{-+}(\omega,\qv)$. Altogether, these observations show that the symmetry and long-wavelength dispersion of magnetic excitations are powerful tools for the characterization of magnetically ordered electronic phases, especially when the complexity of the unit cell increases.

Motivated by these observations, our main question here is what kind of magnons appear in correlated electron systems with unusual magnetic order (i.e., more complicated than a simple ferromagnet or N\'{e}el antiferromagnet). For this purpose, we wish to have a simple model that has tunable magnetic properties while having few model parameters. The Lieb-kagome family of lattices meets these criteria, since the Lieb limit is a ferrimagnet and therefore should have minority magnons, whereas the kagome regime has strong magnetic frustration which favors phases with zero net magnetic moment. 

To be able to answer our main question, we need a computational method that can determine the dynamic spin susceptibility $\chi(\omega,\qv)$ with a high momentum and energy resolution at a feasible computational cost. Few such methods exist, the combination of Hartree-Fock (HF) and Random Phase Approximation (RPA) is one example. It does have the disadvantage that it treats (magnetic) fluctuations in a very crude way and overestimates the tendency towards symmetry breaking. Still, it is often the method of choice in exploratory studies of new magnetic phenomena~\cite{MaierOkamoto23,PhysRevB.110.144412}. We use additional dynamical mean-field theory (DMFT) and exact diagonalization (ED) results as complementary checks on the magnetic ordering tendencies.

The momentum resolution is determined by the linear size $L$ of the system, while the total number of momenta in the calculation scales as $L^d$, so choosing a two-dimensional instead of three-dimensional lattice enables a high resolution at low cost. At the HF level, and even in DMFT~\cite{schafer2015fate}, the phase diagram is only weakly influenced by the dimensionality of the system. In this sense, the magnon results presented here are meant to qualitatively capture ferrimagnets in general rather than quantitatively describing the two-dimensional physics in particular.

The paper is structured as follows: We first introduce the model and the HF + RPA method (Sec.~\ref{sec:model}) and discuss the implications of the Lieb and Mermin-Wagner theorem for our results (Sec.~\ref{sec:theorems}), followed by the Lieb-kagome lattice phase diagram (Sec.~\ref{sec:results:phasediagram}), and finally the magnon spectra in different magnetic phases (Secs.~\ref{sec:symmetry}-\ref{sec:higgs_mechanism}).

\section{Model and Method}
\subsection{Hubbard Model: Lieb and kagome}
\label{sec:model}

\begin{figure*}
\includegraphics{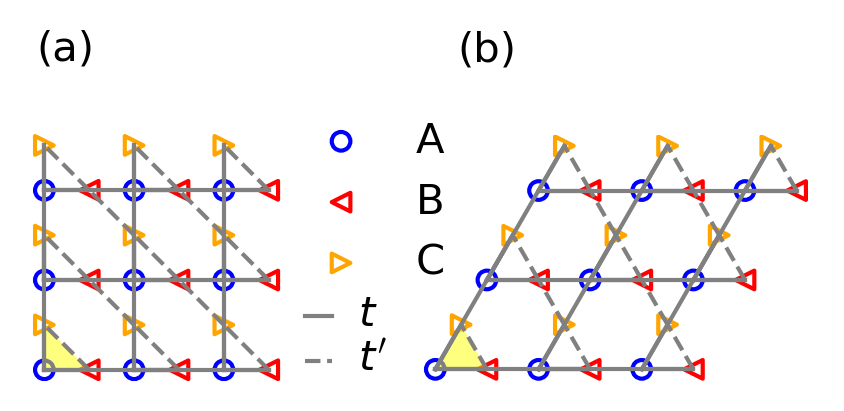}
\raisebox{-0.4cm}{\includegraphics{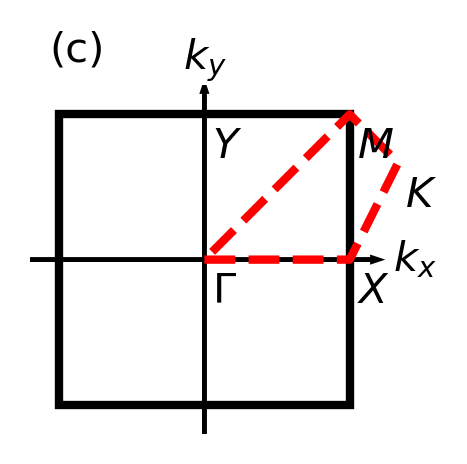}}
\raisebox{-1.0cm}{\includegraphics{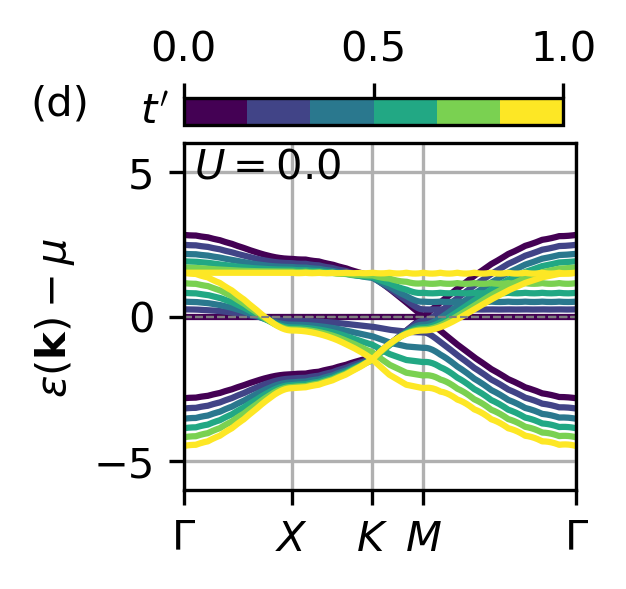}}
\caption{Schematic of (a) the Lieb and (b) kagome lattices. Sites of sublattices A, B, and C are marked by blue circles, red triangles, and orange triangles, respectively, and the shaded yellow region highlights the three sites (A, B, and C) in a single unit cell. Solid and dashed lines refer to the nearest-neighbour and the next-nearest-neighbour hopping amplitudes $t$ and $t'$, respectively. By varying $t'$, the system is continuously tuned from the ideal Lieb ($t'=0$) to the ideal kagome ($t'=t$) lattice. 
(c) The first Brillouin zone and high-symmetry points corresponding to the Lieb lattice.
(d) Non-interacting band structures of the Lieb lattice as functions of $t'$, plotted along the high-symmetry path in panel (c).
}
\label{fig:LiebtoKagome}
\end{figure*}

In the context of the Hubbard model, the lattice structure is reflected in the number of orbitals per unit cell and the tight-binding parameters. Mathematically, the tight-binding parameters can be represented as a weighted graph, and it is only the structure of this graph that matters for the model, not its geometrical interpretation. As discussed by Jiang et al.~\cite{jiang2019topological}, in this way, it is possible to continuously interpolate between the Lieb lattice and the kagome lattice in two dimensions by changing the weights in the graphs, i.e., the hopping parameters.

In both lattices, each unit cell contains three sites. For the Lieb lattice, these are a corner site labeled A and two edge-centered sites labeled B and C, as illustrated in Fig. \ref{fig:LiebtoKagome}(a). In our model, the hopping between the corner site and an edge-centered site is $t$, while the hopping between two edge-centered sites in the same unit cell is $t^\prime$. The situation $t'=0$ corresponds to the normal Lieb lattice, while $t^\prime=t$ gives the kagome lattice, shown in Fig.~\ref{fig:LiebtoKagome}(b), so by varying the parameter $t^\prime\in [0,t]$, we can continuously go from Lieb to kagome. The corner site and the edge sites are inequivalent except at $t'=t$. The Lieb lattice at $t'=0$ has the $C_{4v}$ point group while at finite $t'$, the point group reduces to $C_{2v}$ with symmetry operations $E,~C_{2z},~\sigma_{d},~\sigma_{\bar{d}}$ plus all lattice translations. We set $t=1$ as the unit of energy.
Since the lattices in Fig.~\ref{fig:LiebtoKagome}(a-b) are equivalent (weighted) graphs, either one can be used for the implementation. In our implementation, we use the lattice of Fig.~\ref{fig:LiebtoKagome}(a), i.e., a square lattice with three atoms in the unit cell.

In momentum space, the corresponding Brillouin zone is also square, as shown in Fig~\ref{fig:LiebtoKagome}(c), with high-symmetry points $\Gamma=(0,0)$, X$=(\pi,0)$, $K=(4\pi/3,2\pi/3)$ and $M=(\pi,\pi)$. Note that, in contrast to the square lattice with 1 orbital per cell, here the unit cell with finite $t'$ itself lacks $C_{4z}$ rotation symmetry, which is reflected in a reduced symmetry in the Brillouin zone as well. We use the path $\Gamma$-$X$-$K$-$M$ in the Brillouin zone to visualize the bands. The band structure of the non-interacting model is shown in Fig~\ref{fig:LiebtoKagome}(d) and discussed in App.~\ref{app:noninteracting}; it has a flat band precisely at $t'=0$ and $t'=1$, but not anywhere in between~\cite{jiang2019topological}. Still, the reduced bandwidth of one of the bands can lead to relatively sharp magnetic excitations~\cite{Skovhus24}.

With this definition of the tight-binding model, the Lieb-kagome Hubbard Hamiltonian is given by
\begin{eqnarray}
H=&&-t\sum_{\langle i,j\rangle, \sigma}(c^\dagger_{i\sigma}c_{j\sigma}+\text{h.c.})\nonumber\\
&&-t^\prime\sum_{\langle\langle i,j\rangle\rangle, \sigma}(c^\dagger_{i\sigma}c_{j\sigma}+\text{h.c.})\nonumber\\
&&+U\sum_{i}n_{i\uparrow}n_{i\downarrow}-\mu\sum_i(n_{i\uparrow}+n_{i\downarrow}),
\end{eqnarray}
where $c^\dagger_{i\sigma}$ and $c_{i\sigma}$ are the creation and annihilation operators for an electron with spin $\sigma$ at lattice site $i$, and $n_{i\sigma}=c^\dagger_{i\sigma}c_{i\sigma}$ is the number operator. 
The chemical potential $\mu$ controls the electron density, which is set to half-filling in this study, and $U$ is the on-site Hubbard interaction.

The magnetic ground states of the model are solved at finite temperature using the HF approximation and DMFT. 
We only consider solutions that retain the unit-cell translation symmetry, in the sense that the Hartree-Fock single-particle density matrix $\rho_{i\sigma, j\sigma'} \equiv \langle c^\dagger_{i\sigma} c_{j\sigma'}\rangle$, where  $i,j\in \{A,B,C\}$ and $\sigma,\sigma'=\uparrow,\downarrow$, is the same in every unit cell. For DMFT, we have a similar condition on the Green's function. The HF self-consistency loop is seeded with tiny magnetic fields in the longitudinal and transverse directions to ensure that symmetry-broken solutions, i.e., the ferromagnetic or ferrimagnetic states, and magnetic frustration of the Lieb-kagome lattice can be found. 
To look specifically for altermagnetism, we also applied a small staggered field $S^z_B-S^z_C$ at the first HF iteration. 

The lattice susceptibility $\chi(\omega,\mathbf{q})$ is subsequently calculated from the HF solution, which provides the results with a high momentum and energy resolution and without analytical continuation. 
Starting from the (lesser and greater) single-particle Green's function in real time and momentum space,
\(
G^<(t, \mathbf{k}) = i \rho \, e^{-i \epsilon^{\text{HF}}_{\mathbf{k}}  t}  \, ,~
G^>(t, \mathbf{k}) = i (\rho - \mathbf{1})  e^{-i \epsilon^{\text{HF}}_{\mathbf{k}} t} 
\, ,
\)
where $\epsilon^{\text{HF}}_\mathbf{k}$ is the Hartree-Fock dispersion, and using the spatial Fourier transform $G^\lessgtr(t, \mathbf{r})=\mathcal{F}_{\mathbf{k}\rightarrow\mathbf{r}} G^\lessgtr(t, \mathbf{k})$, the bare lattice susceptibility $\chi_0$ is calculated as the direct product
\begin{multline}
    \chi_0(t, \mathbf{r}) = 
    \\
    i G^<(t, \mathbf{r}) G^>(-t, -\mathbf{r}) 
    - i G^>(t, \mathbf{r}) G^<(-t, -\mathbf{r})
    \, ,
\end{multline}
and subsequently Fourier transformed back to real frequency and momentum
\begin{equation}
    \chi_0(\omega,\mathbf{q})=\mathcal{F}_{\{t,\mathbf{r}\}\rightarrow\{\omega,\mathbf{q}\}}\{
    \chi_0(t, \mathbf{r})
    \}.
\end{equation}
This approach makes it possible to study the real-frequency dependence on dense momentum grids, without having to resort to analytical continuation. This is essential for simultaneously resolving sharp and broad features in the susceptibility, at low and high energies. The numerical implementation uses the Two-Particle Response Function toolbox (TPRF)~\cite{tprf}, which is part of the TRIQS codebase~\cite{parcollet2015triqs}. Finally, the RPA lattice susceptibility is evaluated using the Bethe-Salpeter equation,
\begin{equation}\label{RPAsusceptibility}
\begin{aligned}
    \chi^{\mathrm{RPA}}&=\chi_0+\chi_0U\chi^{\mathrm{RPA}}
    =\frac{\chi_0}{1-\chi_0U},
\end{aligned}
\end{equation}
which is diagonal in frequency and momentum. The bare ($\chi_0$) and RPA ($\chi^{\mathrm{RPA}}$) susceptibilities and the interaction $U$ are all four-dimensional tensors in the spin-orbital indices. 

We will focus on the imaginary part of the dynamical transverse susceptibility, $\Im\chi^{+-}(\omega,\qv)$, which corresponds to the magnetic excitation spectrum, characterizing the dissipative response of the system~\cite{Skovhus24}.

\subsection{Two theorems: Lieb and Mermin-Wagner}
\label{sec:theorems}

Lieb's theorem~\cite{liebtheorem1989} states that the ground state of an unbalanced, bipartite lattice has a finite total magnetization. Here, bipartite means that the system can be divided into two groups of sublattice sites and that hopping only happens between sites from different groups, and the word ``unbalanced" refers to the difference in the number of sites contained in each group. In our model, Lieb's theorem only holds at $t'=0$, since the lattice is only bipartite there. 
We note that Lieb's theorem applies to the zero-temperature ground state, while our study takes place at finite temperatures.

The Mermin-Wagner theorem states that there is no spontaneous symmetry breaking of continuous symmetries in two-dimensional systems at $T>0$~\cite{merminwagner}. This theorem rules out ferrimagnetism as the exact solution. However, approximate solutions such as HF and even dynamical mean-field theory are known to show ordered solutions at low temperature, in violation of the Mermin-Wagner theorem. Our interest here is in the physics of magnetic excitations in ferrimagnets, which can be studied with this model and method, even though the exact solution of the model is not ferrimagnetic. 
In reality, Mermin-Wagner is enforced by very long-wavelength magnetic fluctuations, so that the local physics is often reasonably similar to what is found in approximate, Mermin-Wagner-violating methods~\cite{schafer2015fate,scholle2023comprehensive, scholle2024spiral}. Furthermore, in real two-dimensional materials, the Mermin-Wagner theorem can be broken due to the presence of a substrate that stabilizes the magnetism or due to magnetic anisotropy~\cite{gibertini2019magnetic} stemming from spin-orbit coupling. 

\section{Results}

\subsection{Phase Diagram}
\label{sec:results:phasediagram}

\begin{figure}
\includegraphics{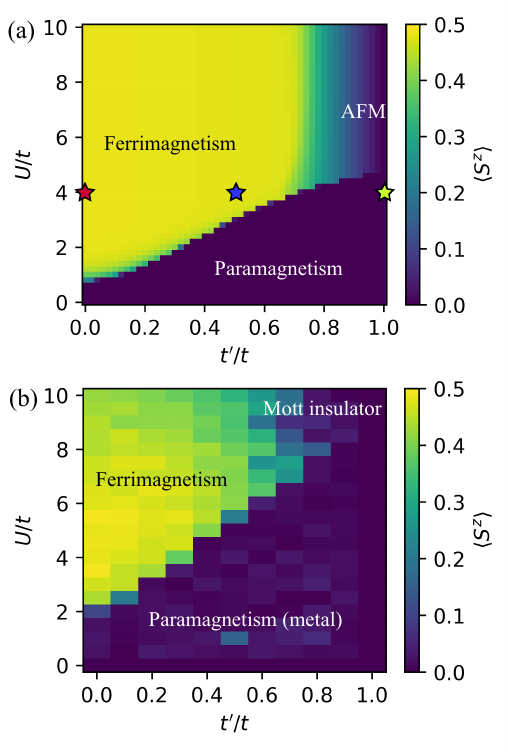}
\caption{(a) Phase diagram in the HF approximation of the Lieb-kagome lattice at the inverse temperature $\beta=10$. The heatmap presents the magnetization as a function of $U$ and $t^\prime$. There is a first-order phase transition from the paramagnetic state to a magnetically ordered state, either ferrimagnetic or antiferromagnetic (AFM), with increasing interaction strength, and a continuous transition between the ferrimagnetic and antiferromagnetic state as a function of $t'$. (b) The corresponding dynamical mean-field theory (DMFT) phase diagram at $\beta=10$. DMFT gives the same paramagnetic to ferrimagnetic transition at moderate $U$. Due to local quantum fluctuations and correlation effects beyond HF, it shows a non-magnetic Mott insulator at large $U$ where the magnetic order is suppressed. 
}
\label{fig:phaset}
\end{figure}

We present the phase diagram of the Lieb-Kagome lattice in the Hubbard model at half-filling in Fig. \ref{fig:phaset}, in terms of the total magnetization $\sum_{i\in\{A,B,C\}}\av{S^z_i}$ as a function of Hubbard interaction $U$ and the hopping parameter $t^\prime$ at the inverse temperature $\beta=10$, computed by the HF approximation and in DMFT (App.~\ref{sec:DMFTmethod}).
From Fig.~\ref{fig:phaset}, we can (for any fixed $t'/t$) identify the critical interaction $U_P$ that separates the non-magnetically ordered (paramagnetic) phase and the magnetically ordered phases. The transition is first order in both HF and DMFT; the magnetization drops to zero as $U<U_P$.

In the Lieb lattice limit ($t'=0$), $U_P$ is found to be relatively small but not zero, as we are at finite temperature. $U_P$ decreases as $U_P\sim1/\beta$ while approaching to the zero-temperature limit, as shown in Appendix~\ref{app:temperature}.
In the HF approximation, this behavior originates from the presence of a flat band at the Fermi energy in the Lieb lattice, where even a small repulsive interaction $U$ can induce a ferrimagnetic ground state at half-filling. Indeed, the Stoner criterion $U D(E_F)\leq 1$ implies a vanishing critical $U$ at $T=0$ due to the divergent density of states, which becomes non-zero when finite temperature is considered.

As $t'$ increases, the flat band acquires dispersion, and the density of states at the Fermi level is gradually reduced. Consequently, both the local-moment threshold and the ordering threshold shift to larger $U$, increasing steadily with $t'$ up to $t'/t\sim0.8$. Both HF and DMFT capture this trend, although the critical interaction in DMFT is systematically larger, $U_P^{\rm HF}<U_P^{\rm DMFT}$, for any given $t'$. This is also supported by finite-cluster exact diagonalization (ED) through ferrimagnetic correlation functions (App.~\ref{sec:EDmethod}), which shows a similar behaviour as DMFT, as discussed in App.~\ref{app:phasediagram}.
This is because the mean-field nature of HF overestimates the tendency to order by neglecting local quantum fluctuations.

\begin{figure*}[ht!]
    \centering
        \includegraphics[width=0.32\textwidth]{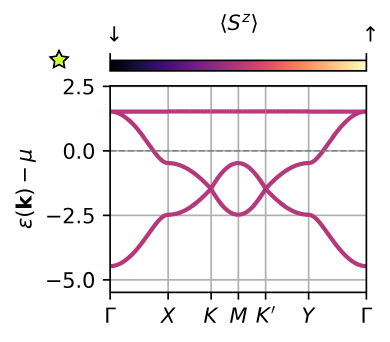}
    \includegraphics[width=0.32\textwidth]{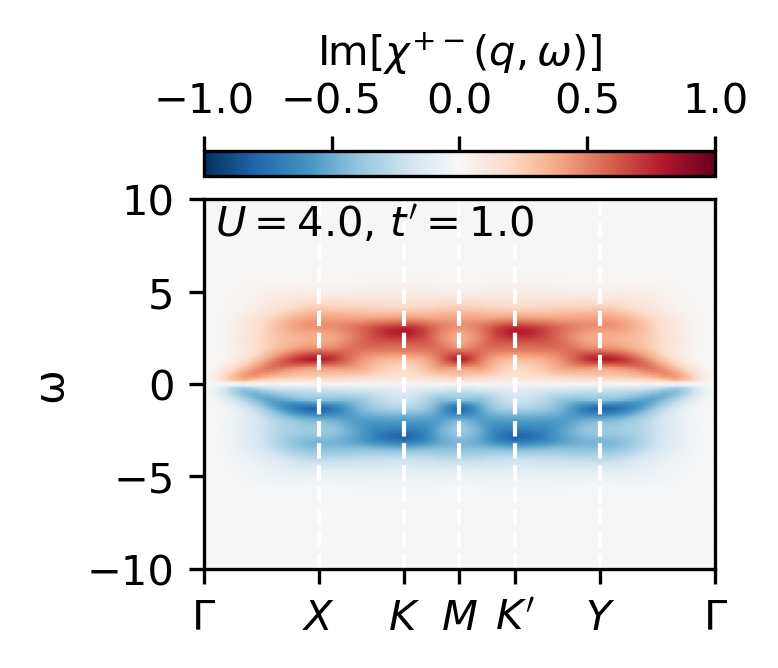}
    \includegraphics[width=0.32\textwidth]{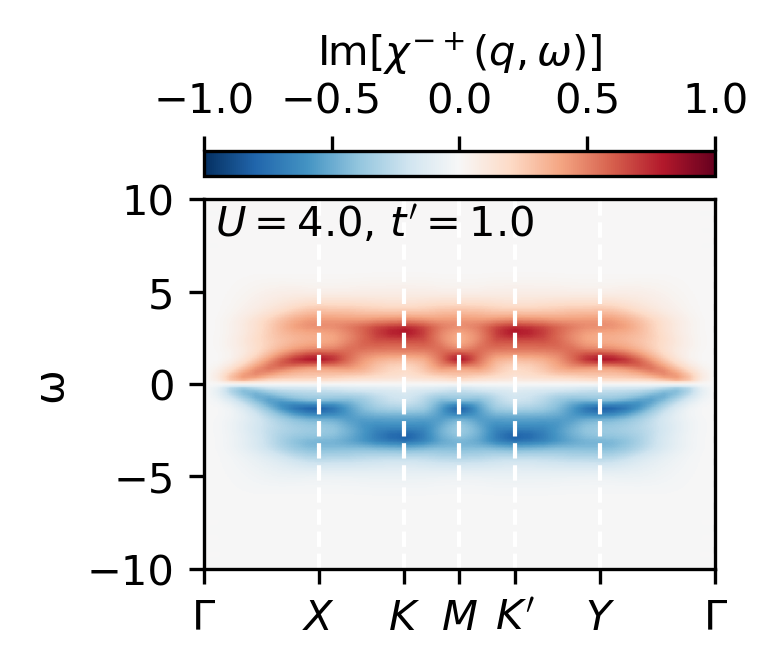}\\
    \includegraphics[width=0.32\textwidth]{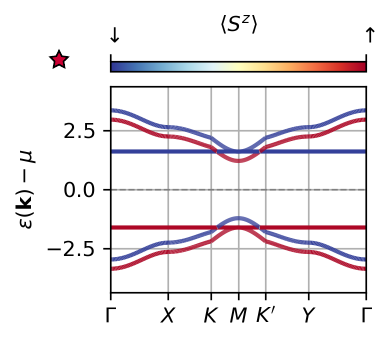}
    \includegraphics[width=0.32\textwidth]{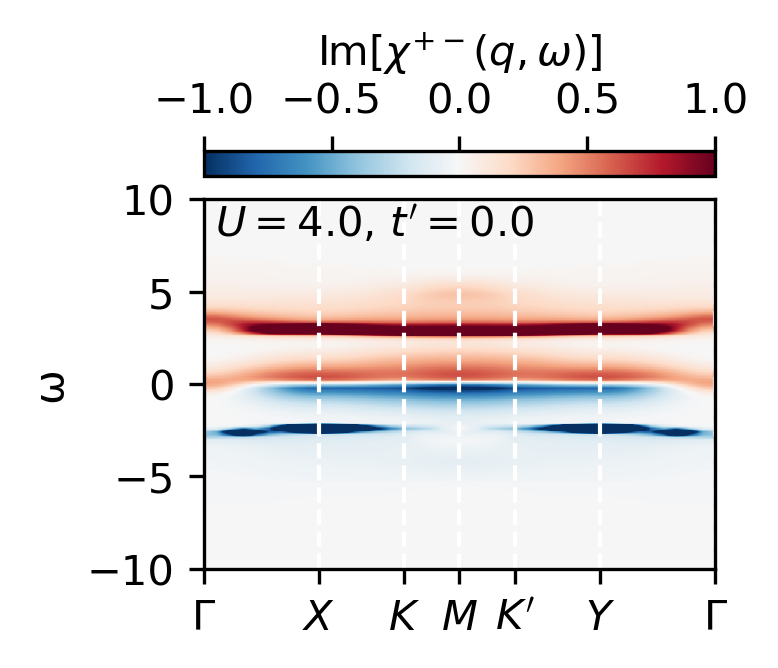}
    \includegraphics[width=0.32\textwidth]{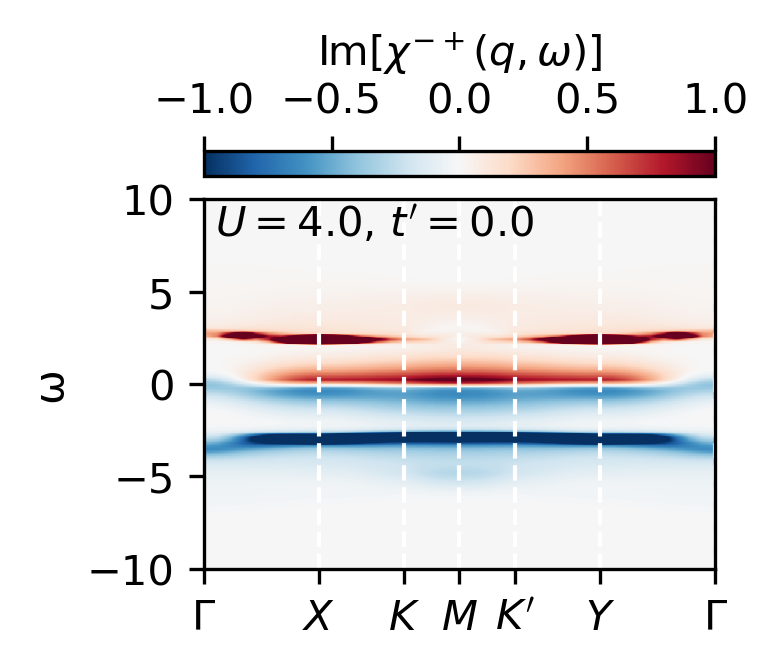}\\
    \includegraphics[width=0.32\textwidth]{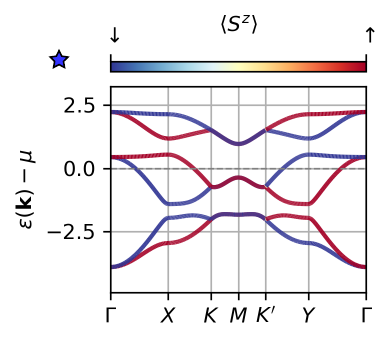}
    \includegraphics[width=0.32\textwidth]{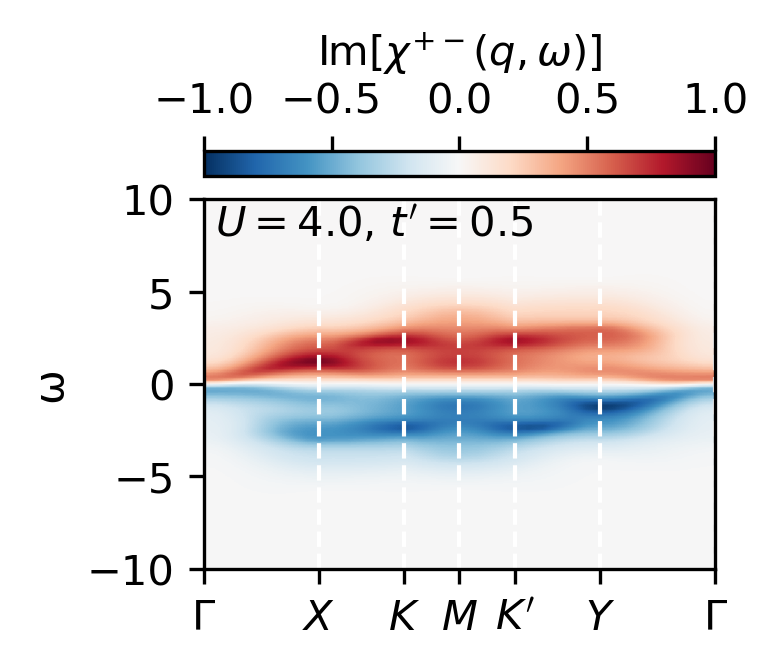}
    \includegraphics[width=0.32\textwidth]{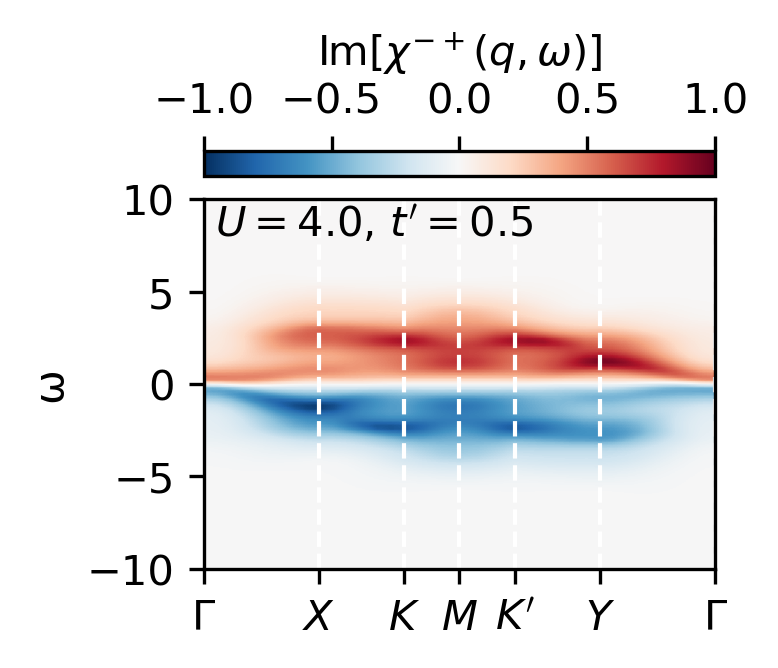}
    \caption{Hartree-Fock band structures and transverse spin susceptibilities $\chi^{+-}$ and $\chi^{-+}$ of three typical magnetic phases in the Lieb-kagome lattice with $U=4.0$. Top row: The paramagnetic state with $t'=1.0$, the kagome lattice limit (marked by the green star in Fig.~\ref{fig:phaset}(a)); Middle row: The ferrimagnetic state with $t'=0.0$, the Lieb lattice limit (marked by the red star in Fig.~\ref{fig:phaset}(a));
    Bottom row: The altermagnetic state with $t'=0.5$ (marked by the blue star in Fig.~\ref{fig:phaset}(a), with applying a specific stagger initial guess).} 
    \label{fig:symmetry}
\end{figure*}

Going towards the kagome lattice ($t'/t\to1$), magnetic frustration becomes important, and a continuous transition to an antiferromagnetic state on the corner-sharing triangles is found in HF. This antiferromagnetic state only occurs for a narrow range of $t'$ close to the kagome limit. However, it is important to note that the HF approximation employed here neglects correlation effects beyond the mean field and generally overestimates the stability of symmetry-broken states and the magnetic order~\cite{vanLoon2023larmor}. As a result, it cannot fully capture the frustration in the kagome lattice accurately. Therefore, the critical $t'$ at a certain $U$ for the transition of antiferromagnetic order is larger than that reported in other studies using more sophisticated methods \cite{lima2023magnetism, yamada2011mott}. Indeed, already in DMFT, the $t'/t\rightarrow 1$ regime shows a nonmagnetic Mott insulator instead. 

In the strongly correlated regime, the DMFT solution can be viewed as an effective antiferromagnetic Heisenberg model with an ordering temperature $T_N^d\sim J\sim 1/U$~\cite{kumar2017Liebdome}, so that at sufficiently low but finite temperature, the magnetism may be suppressed. In this sense, DMFT captures an important reduction of the ferrimagnetic order parameter at large $U$, visible in Fig.~\ref{fig:phaset}, similar to the reduction of antiferromagnetic order in DMFT for the square or cubic lattice~\cite{schafer2015fate,PhysRevB.92.144409}. This effect is absent in HF, which remains magnetic in regimes where DMFT already predicts a Mott-insulating state.

Overall, DMFT provides the same ordering trends at moderate interaction strength in Fig.~\ref{fig:phaset}, but it goes beyond HF by incorporating local dynamical correlations. The comparison between HF and DMFT therefore shows that HF is qualitatively reliable in much of the phase diagram, especially for identifying the different magnetic phases and the overall shape of the phase boundaries.
A detailed discussion on the Lieb-kagome lattice phase diagram is given in App.~\ref{app:phasediagram}.

\subsection{Symmetries of Transverse Susceptibility}
\label{sec:symmetry}

The different magnetic phases have characteristic signatures in the electronic band structure and in the spin susceptibility. Three representative examples are shown in Fig.~\ref{fig:symmetry}. 

The top row of Fig.~\ref{fig:symmetry} shows a paramagnetic solution without splitting of the electronic bands. There is no spin polarization and $SU(2)$ symmetry is not broken. In this case, the transverse magnetic susceptibility is characterized by the symmetry
\begin{equation}
   \Im \chi^{+-}(\omega, \mathbf{q})= \Im\chi^{-+}(\omega,\mathbf{q}),~\rm{(paramagnet).}
\end{equation}
The susceptibility close to the $\Gamma$ point shows the characteristic paramagnon modes with vanishing energy as $\qv\rightarrow 0$.

In contrast, the middle row of Fig.~\ref{fig:symmetry} shows an insulating ferrimagnetic solution at $t'=0$ (Lieb lattice). In this case, there is an exchange splitting between $\uparrow$ and $\downarrow$ bands in the electronic structure, which leads to two filled majority bands and one filled minority band and therefore to a net magnetization. Both spin flavors still show the characteristic flat band of the Lieb lattice. The breaking of $SU(2)$ symmetry leads to differences between $\chi^{+-}$ and $\chi^{-+}$, but we still have the relation
\begin{equation}
    \Im\chi^{+-}(\omega, \mathbf{q})= -\Im\chi^{-+}(-\omega,\mathbf{q}),~\rm{(ferrimagnet).}
\end{equation}
For both channels, rotation symmetry within the Brillouin zone, $\Im\chi^{+-/-+}(\omega,(q_x,q_y))=\Im\chi^{+-/-+}(\omega,(q_y,q_x))$, holds because of the equivalence of the sites B and C.
Noticeably, we can find a gapped, nearly flat magnon band in both majority and minority channels in the symmetry-broken ferrimagnetic phase. We refer to this as the Higgs (amplitude) magnon mode~\cite{pekker2015amplitude}, which will be further discussed in the subsequent sections.

More interestingly, when the system is driven into the altermagnetic phase by applying a staggered initial field that favors altermagnetic order, the spectra display a different symmetry, as shown in the bottom row of Fig.~\ref{fig:symmetry} with $U=4.0$ and $t'=0.5$ as an example. For the band structure, spin splitting is visible, together with the fact that the splitting is inverted when performing a rotation in the Brillouin zone. In this case, for the magnetic susceptibility, we find that
\begin{equation}
   \Im \chi^{+-}(\omega,(q_x,q_y))=\Im\chi^{-+}(\omega,(q_y,q_x)),~\rm{(altermagnet).}
\end{equation}
This reflects the characteristic combination of rotational and spin-flipped symmetry of the altermagnetic state~\cite{Smejkal23,MaierOkamoto23,Liu24}, showing the magnonic chirality in the altermagnets.

Previous unrestricted HF studies have shown that the altermagnetic Mott insulator is the ground state of the Lieb lattice only at fillings $n=2$ and $n=4$~\cite{kaushal2025altermagnetism}; modified hopping can instead stabilize an altermagnetic metal~\cite{durrnagel2025altermagneticlieb}.
Consistently, we find that at half-filling ($n=3$) the altermagnetic state obtained in our calculations is metastable, rather than constituting a global ground state within the HF framework. 
In textbook band theory, systems with an odd number of electrons famously have to be metallic due to the spin degeneracy of two of every band. Here, similarly, the paramagnetic and altermagnetic phases have to be metallic since we have an odd number of electrons (3 per unit cell) and both phases have bands with multiplicity two, albeit after performing a rotation in the case of the altermagnet, and thus have a spin-balanced density of states. The ferrimagnet, on the other hand, breaks the spin symmetry in the band structure and density of states, and thus allows for a spin-imbalanced insulating state. 
Since correlated electron systems at low temperature have a strong tendency to avoid metallicity, this explains why it is hard to stabilize altermagnetism at $n=3$ and easier at $n=2$ and $n=4$~\cite{kaushal2025altermagnetism}.
Nevertheless, the corresponding band structure and magnetic excitation spectra can still yield insights into altermagnetic behavior in other materials~\cite{kqwx-v6jv}.

\subsection{Transverse Susceptibility Evolution}
\label{sec:results:susc}

\begin{figure}
\includegraphics[width=0.235\textwidth]{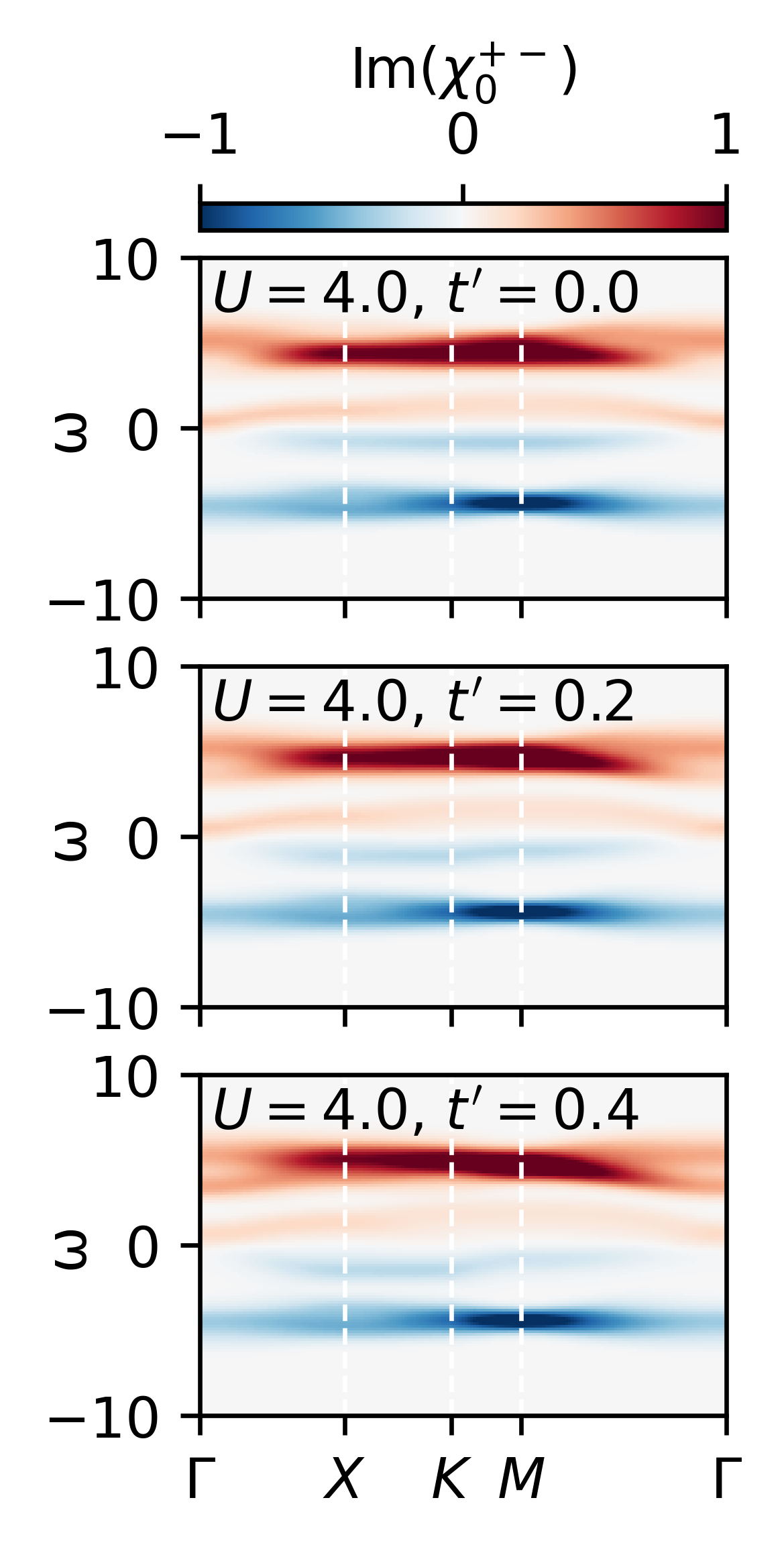}
\includegraphics[width=0.235\textwidth]{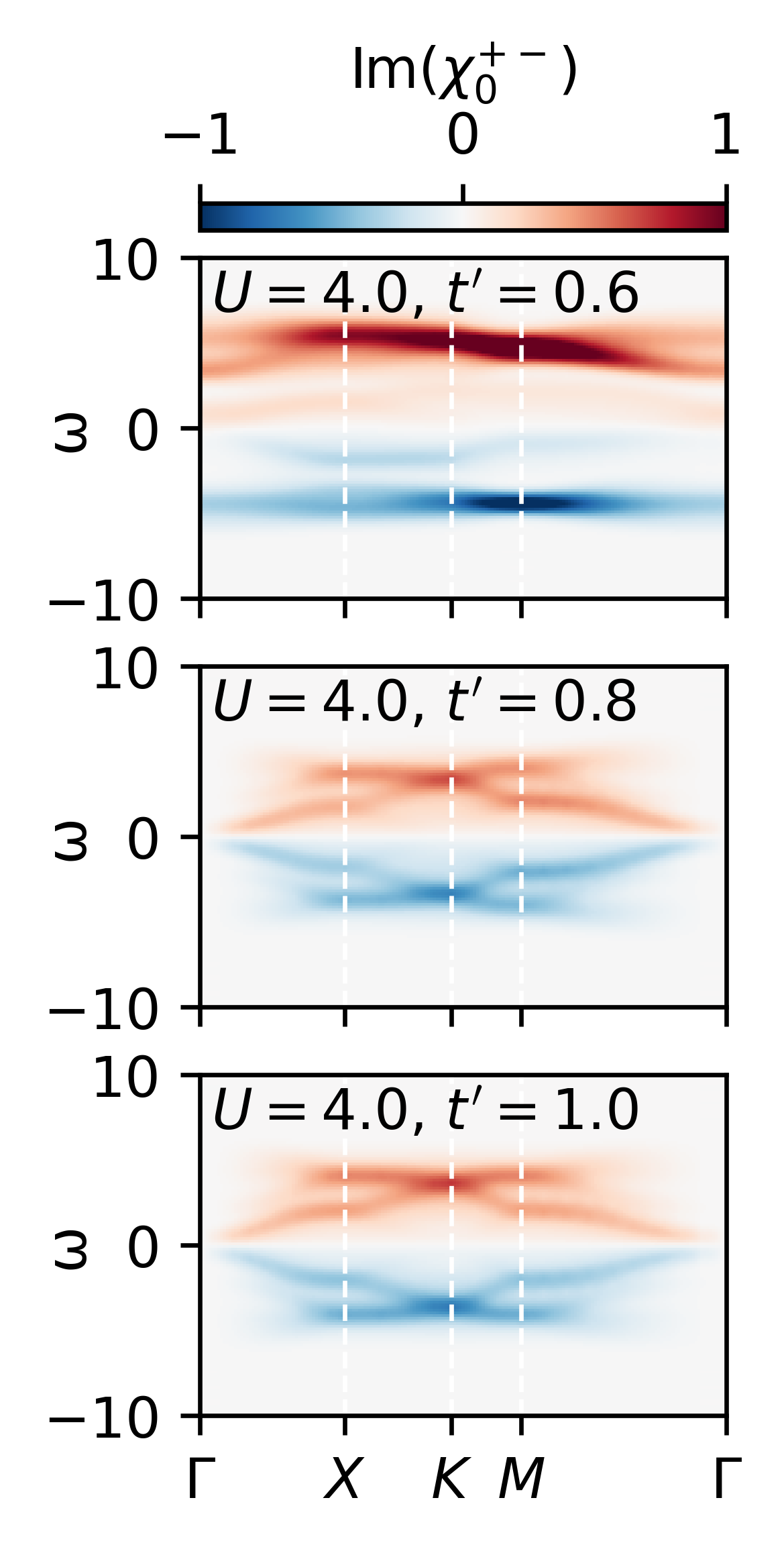}
\caption{Imaginary part of bare magnetic susceptibility $\Im\chi^{+-}_0(\omega,\qv)$ as a function of $t'$ with $U=4$. The ferrimagnetic-paramagnetic transition takes place between $t'=0.6$ and $t'=0.8$.
}
\label{fig:baresusceptibility4.0}
\end{figure}

\begin{figure}
\includegraphics[width=0.235\textwidth]{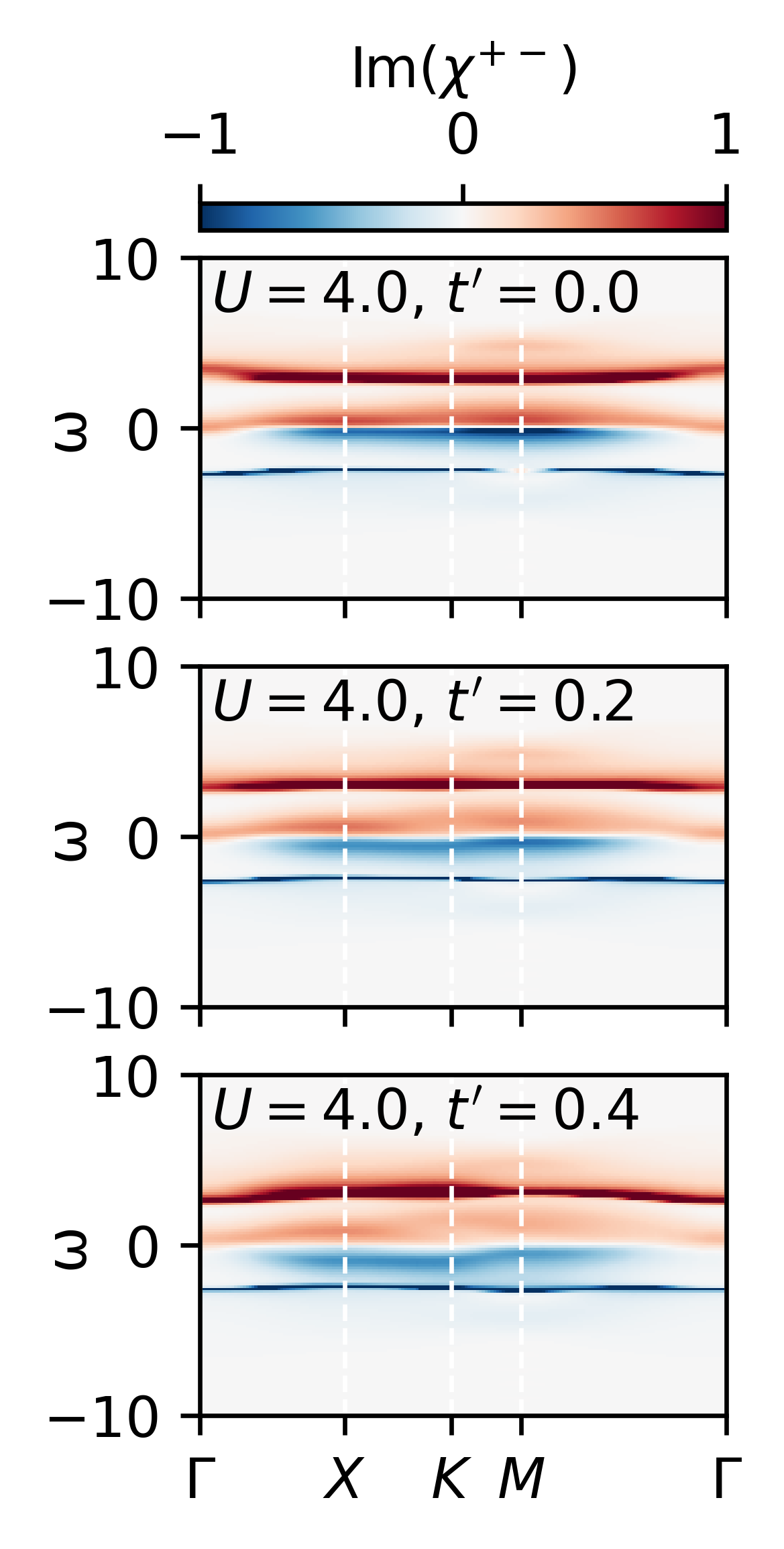}
\includegraphics[width=0.235\textwidth]{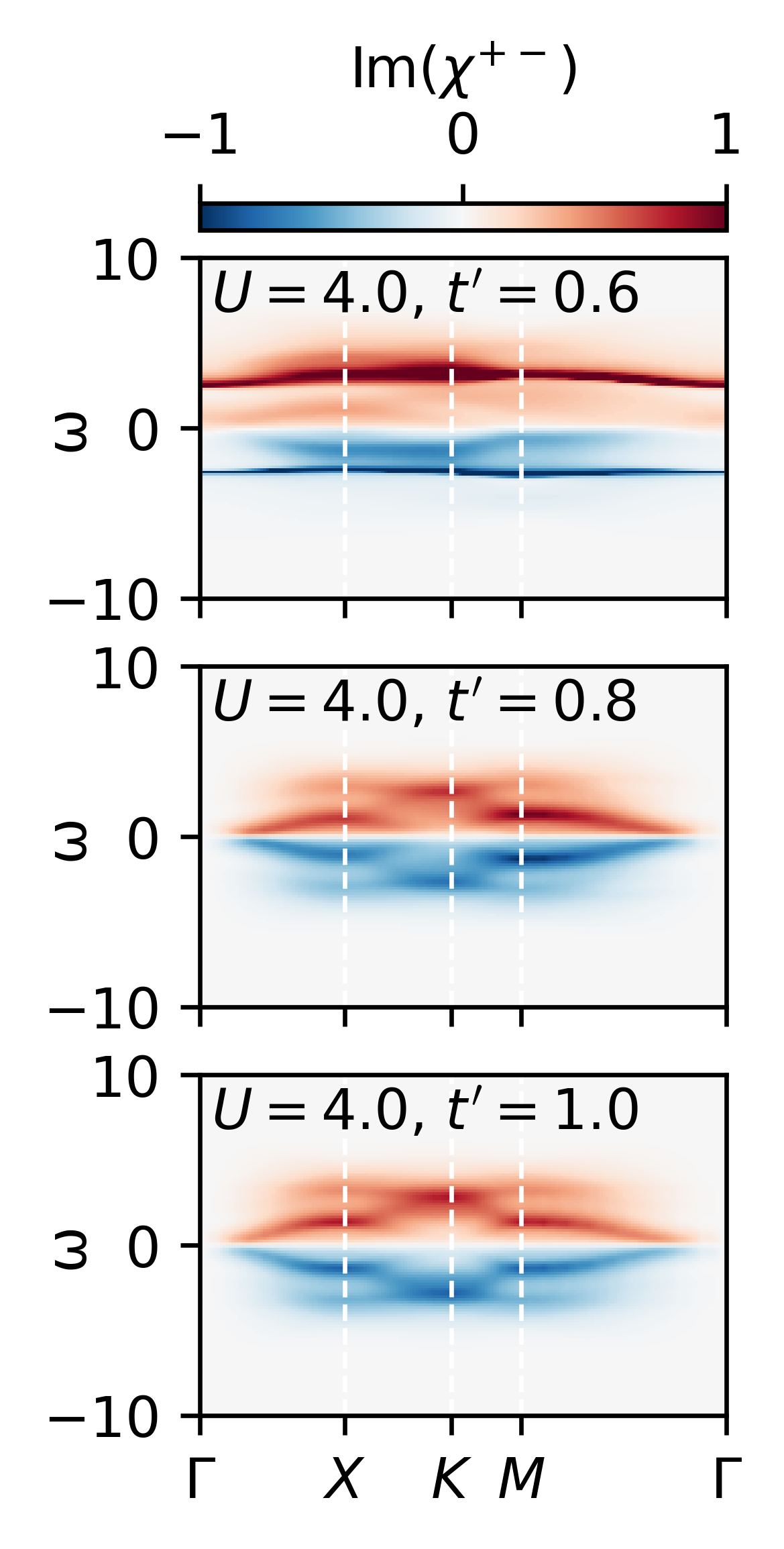}
\caption{Imaginary part of magnetic susceptibility $\Im\chi^{+-}(\omega,\qv)$ computed by RPA as a function of $t'$ with $U=4$. The ferrimagnetic-paramagnetic transition takes place between $t'=0.6$ and $t'=0.8$.
}
\label{fig:susceptibility4.0}
\end{figure}

\begin{figure}
\includegraphics[width=0.5\textwidth]{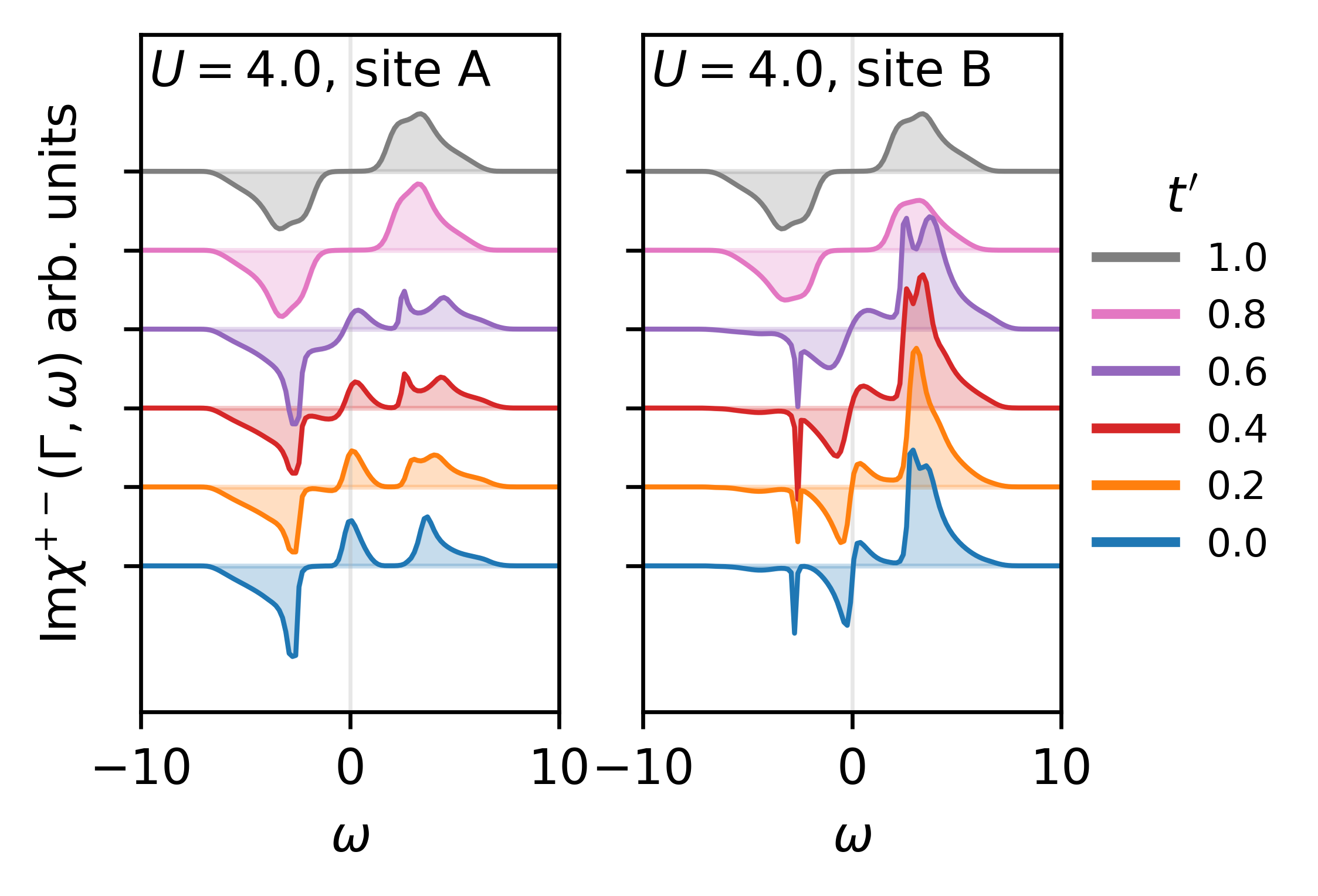}
\caption{Cross-section of the magnetic susceptibility $\Im\chi^{+-}(\Gamma,\omega)$ for interaction strength $U=4.0$, shown separately for site A (left) and site B (right). The comparison highlights how increasing $t'$ shifts and reshapes the peaks of the magnetic susceptibility on the two inequivalent sublattice sites. The sharp peaks (the maximum value is shown with cutoffs in the figure, same as Figs.~\ref{fig:susceptibility:gamma8.0} and ~\ref{fig:susceptibility:gammaU0}) of the Higgs modes disappear in the paramagnetic phase ($t'=0.8$ and $t'=1.0$). }
\label{fig:susceptibility:gamma}
\end{figure}

The imaginary part of the transverse bare lattice susceptibility $\chi_0^{+-}$, computed from the HF band structures at $U=4.0$, is shown in Fig.~\ref{fig:baresusceptibility4.0}. These spectra reflect the magnetic response directly computed from the renormalized single-particle bands. In the ferrimagnetic regime ($t'\le 0.6$), the dominant features are Stoner pairs excitations, which are a broad continuum of single-particle spin-flip transitions. In particular, the strong intensity around $\omega\approx 5$ corresponds to the energy gap between the spin-up and spin-down bands in the ferrimagnetic state. We also observe weaker, but clearly resolved, minority band (anti-Stoner modes) features consistent with the sublattice spin imbalance of the Lieb–kagome lattice. After the transition to the paramagnetic phase ($t'\geq 0.8$), these ferrimagnetic features vanish, and the spectrum becomes featureless at those energies.
Noting that $\chi_0^{+-}$ only includes single-particle (bubble) processes, nevertheless, the collective magnon bands will appear at the level of the RPA.

Figure~\ref{fig:susceptibility4.0} shows the imaginary part of transverse susceptibility $\chi^{+-}$ at $U=4.0$ computed by RPA. 
For small $t'$, the system is in the ferrimagnetic phase, giving the relation $\chi^{+-}(\omega)=-\chi^{-+}(-\omega)$, and the susceptibility shows two rather flat high-energy Higgs modes which correspond to changes in the local magnetic moments. There is a Higgs mode at both positive and negative energy since it is possible to flip a majority spin on the BC sublattice or a minority spin on the A sublattice. Furthermore, there are dispersive magnon (Goldstone) modes at low energy, dispersing down to $\omega=0$ at $\Gamma$. These correspond to orienting the order parameter in another direction, which does not cost energy. Again, there are two of them, since the ferrimagnet has sites with either spin orientation. Deeper inside the Brillouin zone, broadening of the modes takes place due to the dispersion of the electronic bands, which allows for different energies for transitions with the same $\mathbf{q}$. Moving away from $t'=0$, the flat band in the Lieb model becomes more dispersive, and this is reflected in a wider continuum in the susceptibility. 

After the transition to the paramagnetic phase at $t'/t \gtrsim 0.8$ for $U=4.0$, the Higgs modes disappear, since there is no longer an order parameter to change. The Goldstone modes remain present in the form of paramagnons. Furthermore, paramagnetism leads to the equivalence of the positive and negative energy parts of the spectrum, or equivalently to $\chi^{+-}=\chi^{-+}$.

\begin{figure}
    \centering
    \includegraphics[]{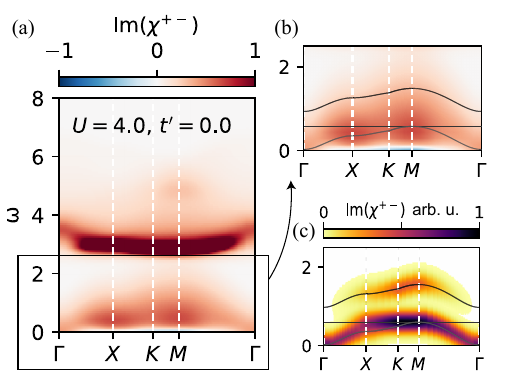}
    \caption{(a) RPA transverse susceptibility, $\Im\chi^{+-}(\omega,\qv)$, for positive $\omega$ only. (b) Zoom on the low-energy sector: the Goldstone (spin-wave) mode magnon spectrum computed with HF+RPA. (c) Spin-wave spectrum and correlation function computed by linear spin wave theory for the corresponding Heisenberg model with $J=4t^2/U=1$; the black lines in (b) and (c) show the spin wave dispersion.}
    \label{fig:spinwaveU4}
\end{figure}

Figure~\ref{fig:susceptibility:gamma} provides additional insight into the Higgs modes by looking at the site-resolved $\Gamma$-point susceptibility. On the minority site A, the minority Higgs mode is visible as the sharp peak at negative energy, which disappears in the paramagnetic phase ($t'=0.8$ and $t'=1$). The energy of the minority Higgs mode is related to the free energy landscape and is rather independent of $t'$. On the majority site $B$, for $0<t'<0.8$, some spectral weight is visible at the same negative frequency. On the other hand, most spectral weight at the B site is present at positive frequency, as expected for a spin-polarized system.

The three panels in Fig.~\ref{fig:spinwaveU4} compare the magnonic spectra computed with the HF+RPA method and the linear spin wave theory (SWT)~\cite{toth2015linear} for the effective antiferromagnetic Heisenberg model, $H=\sum_{\av{i,j}}J_{ij}\mathbf{S}_i\cdot\mathbf{S}_j$ with $J_{ij}=4t_{ij}^2/U$, with the same magnetic order. While the spin-wave theory for the Heisenberg model reproduces the gapless Goldstone modes that match the HF+RPA spectrum in the low-energy window, it fails to capture intrinsically itinerant features visible in the full RPA spectrum at $U=4.0$, $t'=0$, as an example. 
In particular, SWT predicts low-energy, transverse Goldstone magnons but cannot produce the broad, higher-energy Stoner continuum arising from single particle–hole excitations from itinerant electrons, nor the finite-energy amplitude (Higgs-like) peak that appears in the RPA calculation.
The full RPA spectrum, therefore, shows the full Stoner continuum, plus a Higgs mode associated with fluctuations of the order parameter magnitude, both absent in a localized-spin treatment.
Our method identifies and attributes the extra spectral features to itinerant electron physics and collective amplitude fluctuations (at the RPA level), demonstrating the added value of the present approach for a complete description of magnetic excitations.

\begin{figure}
\includegraphics[width=0.235\textwidth]{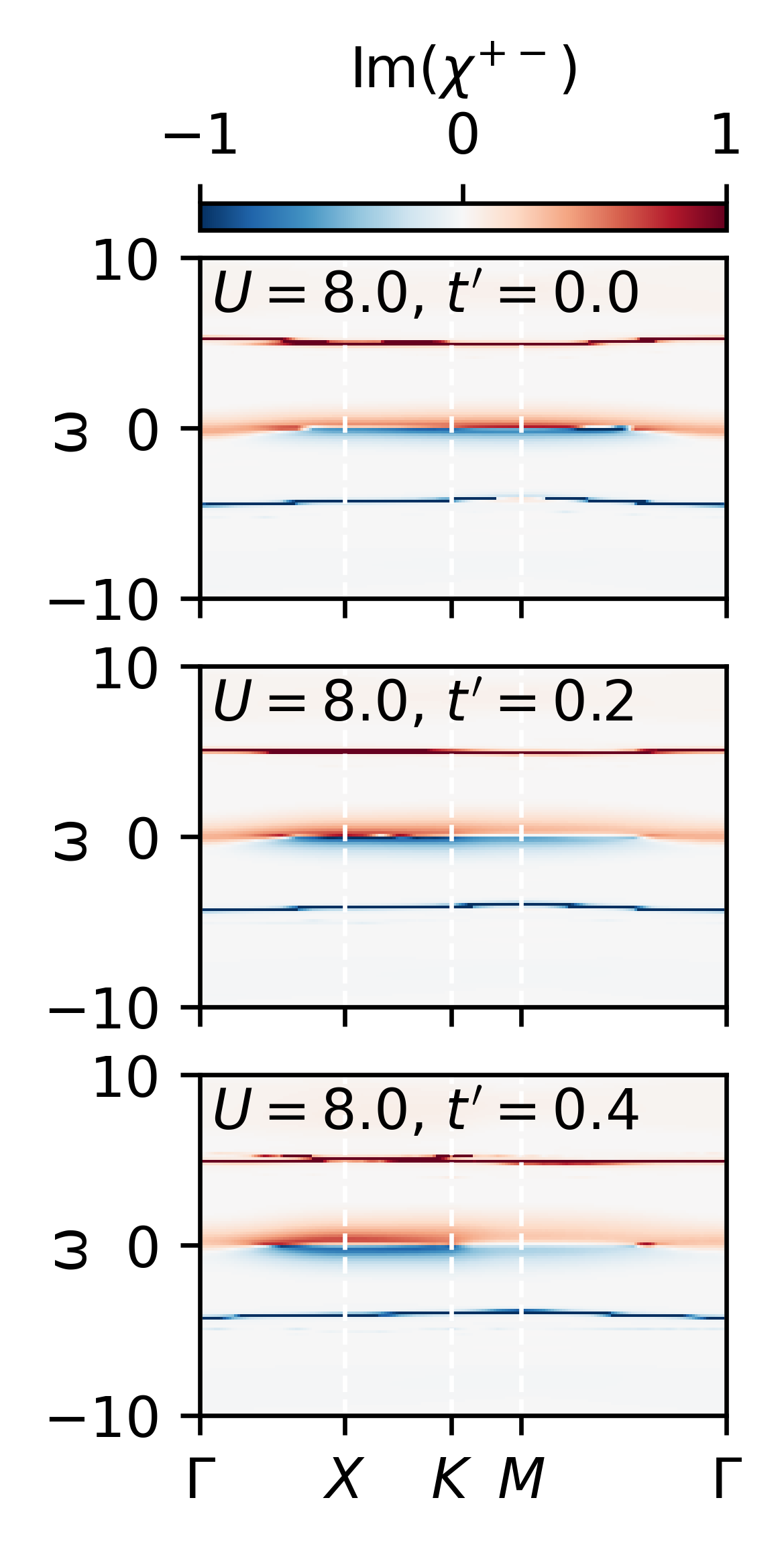}
\includegraphics[width=0.235\textwidth]{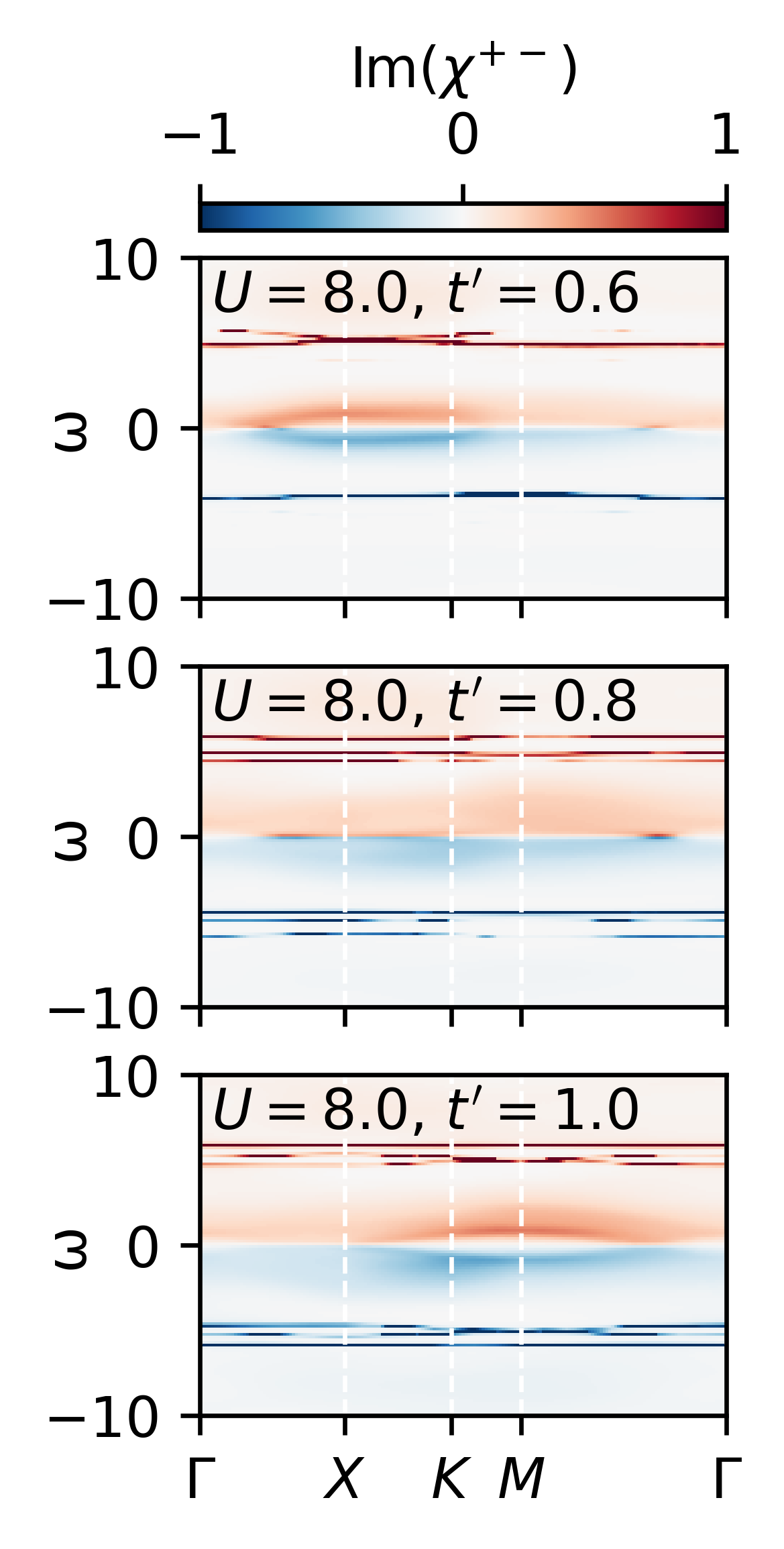}
\caption{Imaginary part of magnetic susceptibility $\Im\chi^{+-}(\omega,\qv)$ computed by RPA as a function of $t'$ with $U=8$. }
\label{fig:susceptibility8.0}
\end{figure}

\begin{figure}
\includegraphics[width=0.5\textwidth]{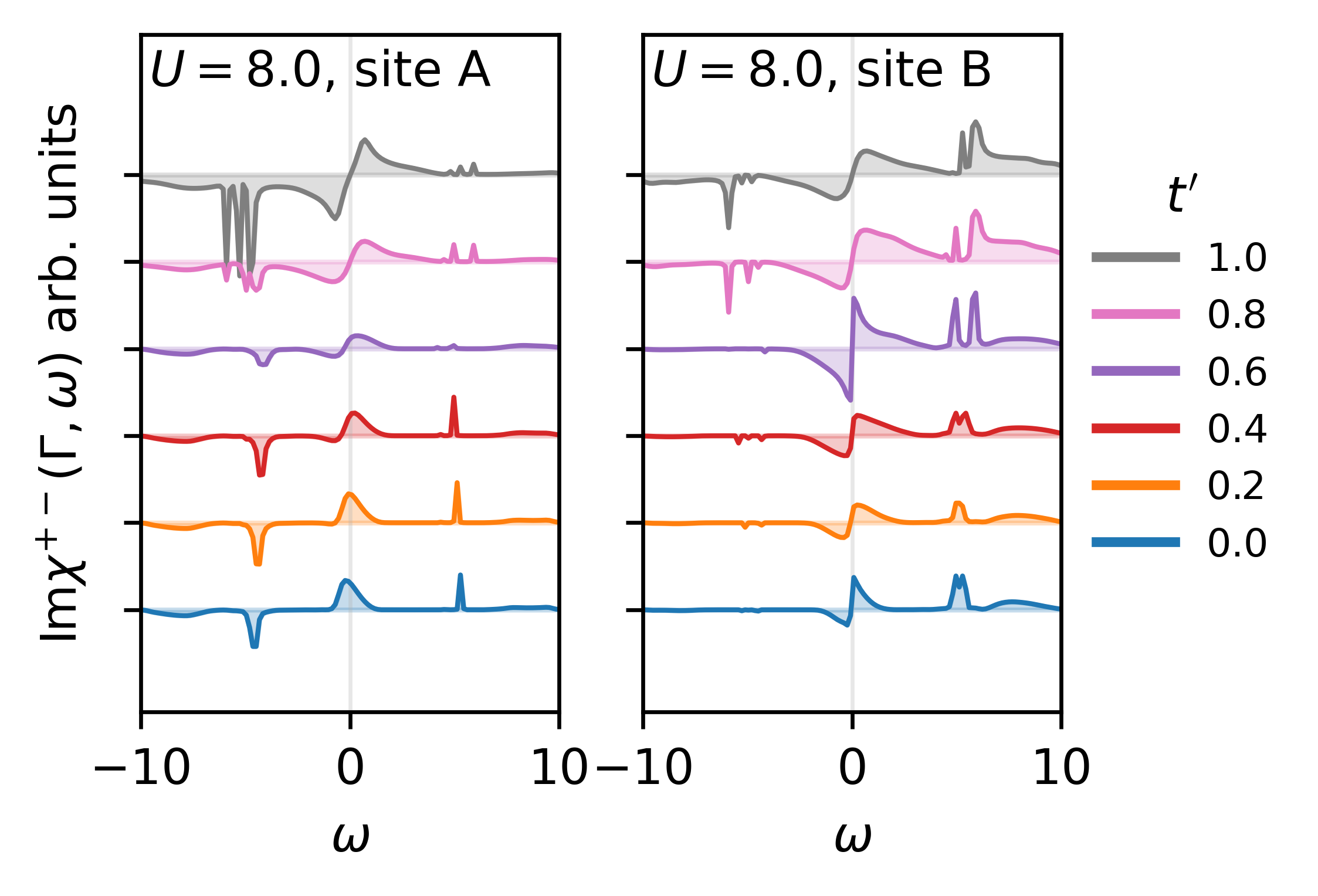}
\caption{Cross-section of the magnetic susceptibility $\Im\chi^{+-}(\Gamma,\omega)$ for interaction strength $U=8.0$, shown separately for site A (left) and site B (right). }
\label{fig:susceptibility:gamma8.0}
\end{figure}

The transverse susceptibility $\chi^{+-}$ for the strong coupling case $U=8.0$ is also plotted in Fig.~\ref{fig:susceptibility8.0}. 
At $U=8.0$, the static HF band splitting is larger than at $U=4.0$, pushing the spin-up and -down bands farther apart. 
As a result, both electron-hole-pair excitation and Higgs mode appear at higher absolute energies in either the majority or minority channels.
Due to the broken symmetry at larger $U$, the Higgs modes still exist with $t'=1$, where the total average magnetization is zero, but the magnetization in the different sublattices is imbalanced in the antiferromagnetic phase.

Additionally, in Fig.~\ref{fig:susceptibility8.0}, one observes a new, weakly dispersing mode just above the original Higgs peak with $t'\gtrsim 0.6$, where the phase transition of ferri- and antiferromagnetism begins to appear.
This additional feature likely reflects an extra Higgs mode that becomes available when the next-nearest-neighbor hopping both broadens the flat band and increases frustration when the system is close to the antiferromagnetic phase, i.e., in the upper right area of the phase diagram.
By contrast, at $U=4.0$ (Fig.~\ref{fig:susceptibility4.0}), the susceptibility maintains only two dominant Higgs peaks throughout the ferrimagnetic phase and disappears in the paramagnetic phase.

The site-resolved $\Gamma$-point susceptibility is shown in Fig.~\ref{fig:susceptibility:gamma8.0} with $U=8.0$ accordingly. Similarly, the minority site A gives the main contribution to the minority Higgs mode, whereas site B dominates the majority mode because of the unbalanced spins.
The Goldstone mode broadens as $t'\rightarrow 1$.
However, different from the case of $U=4.0$, one can observe the double-peak Higgs modes for both majority and minority channels with $t'\geq 0.6$. 
We note that the Higgs modes appear as poles of the RPA susceptibility (Eq.~\ref{RPAsusceptibility}), causing the exact peak heights in our spectra to depend sensitively on numerical choices (frequency grid, padding, etc.) so they should be treated with caution.  The pole positions are the robust physical quantities.

\subsection{Higgs Mode of Magnon}
\label{sec:higgs_mechanism}

\begin{figure*}[htb]
   \includegraphics[width=0.98\textwidth]{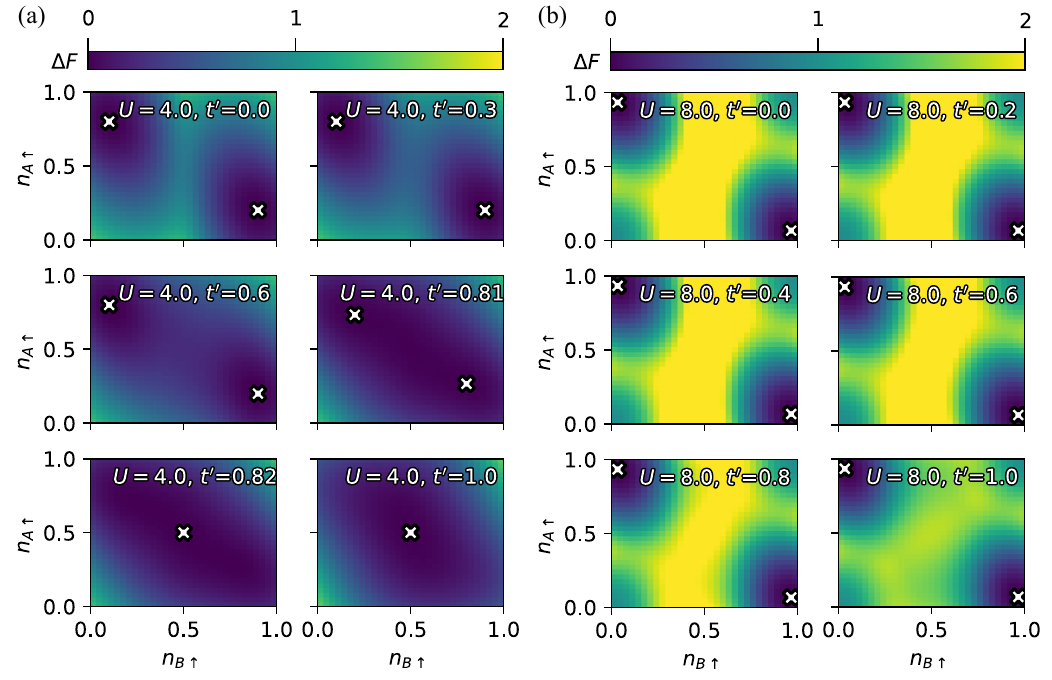} 
   \caption{(a) Helmholtz free energy $F$ landscapes as functions of the number of particles with spin-up at site A ($n_{A\uparrow}$) and site B ($n_{B\uparrow}$) with $U=4$, where the minimum points are denoted by the white crosses. The phase transition to paramagnetism occurs between $t'=0.81$ and $t'=0.82$, where the lowest energy lies in the center with $t'\geq0.82$, indicating no spontaneous symmetry breaking. (b) The same figure as (a) but with $U=8$. The lowest energy at the corner indicates the symmetry-broken ferrimagnetic or antiferromagnetic states.
}
\label{fig:freeenergyU4}
\end{figure*}

The Higgs, clearly observed from Fig.~\ref{fig:susceptibility4.0} and \ref{fig:susceptibility8.0}, is a gapped, longitudinal collective excitation of a magnetically ordered state, which involves oscillations in the magnitude of the order parameter rather than its direction~\cite{pekker2015amplitude}. 
Related gapped modes have also been experimentally observed in ferromagnetic kagome lattice materials~\cite{chisnell2015topological,riberolles2024chiral}.
The curvature of the free-energy minimum sets the Higgs magnon’s energy and therefore becomes higher and better defined when the symmetry-broken minima deepen and separate. 
For this reason, we examine the free-energy map as a function of the order parameters (the site magnetizations), which directly reveals the stiffness of the order parameter and thus provides a clear, intuitive picture of the origin of the gapped Higgs magnon branch.

Figure~\ref{fig:freeenergyU4} shows the Helmholtz free-energy difference, 
$\Delta F(n_{A\uparrow},n_{B\uparrow}) = F(n_{A\uparrow},n_{B\uparrow}) - F_{\mathrm{min}}$,
mapped over the plane spanned by the spin-up occupations on sites A and B (with site C equivalent to B) for various values of $t'$ and interaction strengths $U=4$ and $U=8$.
The comparison between the two interaction strengths shows how the magnetic phase and the degree of symmetry breaking evolve with increasing correlation. For $U=4.0$ (Fig.~\ref{fig:freeenergyU4}(a)), the free-energy landscape exhibits shallow double wells that are close to the central, reflecting a weak sublattice spin imbalance and a small energy gap for amplitude (Higgs) fluctuations. As $t'\rightarrow1$, the minima lie in a single symmetric valley centered at $(n_{A\uparrow},n_{B\uparrow})\!\approx\!(0.5,0.5)$, corresponding to the paramagnetic phase without spontaneous symmetry breaking, consistent with the paramagnetic phase region shown in the phase diagram in Fig.~\ref{fig:phaset}. 
We notice that the critical $t'$ of this phase transition ($t'\approx0.81$ at $U=4$), as examined by the free energy map, is slightly higher than the critical $t'$ in the HF self-consistent calculation ($t'\approx0.75$ at $U=4$). Because the free-energy landscape is very flat near the phase boundary, the self-consistent HF iteration may converge to a local minimum, i.e., the paramagnetic phase near the critical region.
In contrast, for $U=8.0$ (Fig.~\ref{fig:freeenergyU4}(b)), the free-energy minima become deeper and more separated, leading to a stable symmetry-broken magnetic ground state. The increased curvature around the minima implies a stiffer order parameter and a higher-energy Higgs mode.

\section{Conclusion}

To summarize, we investigated how unconventional magnetic order in a correlated multiorbital system is reflected in its magnetic excitation spectrum, using a family of Hubbard models that interpolates between the Lieb and kagome lattices. Going from Lieb to kagome, the magnetic frustration increases and the preferred ordering goes from ferrimagnetic to antiferromagnetic. Goldstone modes corresponding to the free rotation of the order parameter are present in the long-wavelength magnetic susceptibility across the phase diagram. In the ferrimagnetic phase, there are distinct minority and majority Goldstone modes, due to the presence of both minority and majority sites. Compared to the material-specific observation of minority magnons in Fe$_3$GeTe$_2$~\cite{Skovhus24}, the present set-up provides a minimal model for minority magnons with a clear real-space separation of majority and minority electrons. We also find weakly dispersing Higgs magnon modes in the symmetry-broken phases, with an energy that reflects the stiffness of the free energy landscape around the minimum. The multi-orbital character of the model is necessary to get distinct types of magnetic fluctuations at $\qv=0$, similar to the acoustic and optical modes in lattice dynamics. The site-resolved susceptibility further reveals that the Higgs modes can be distinguished on different sublattices, highlighting their symmetry-breaking origin and spin-selective character. 

Our results are mainly based on HF and RPA, supplemented with DMFT and ED as a complementary check of the trends in the magnetic phase diagram. Both HF and DMFT are mean-field theories that tend to overestimate the magnetic ordering, especially in low-dimensional systems, due to the lack of spatial magnetic fluctuations. Still, the mean-field picture enables the qualitative identification of Goldstone and Higgs branches in the magnetic susceptibility as robust indicators of magnetically ordered phases when they do occur.
These magnonic spectra, especially the coexistence of dispersive Goldstone magnons and flat Higgs modes, should be observable in inelastic neutron scattering (INS) or resonant X-ray scattering (RIXS)~\cite{musshoff2021susceptibility}, both in Lieb- or kagome-based materials and in other lattice systems with spontaneously broken symmetry. 
Our findings highlight the connection between magnetic order and magnetic excitation spectra, including the spectral symmetries, sublattice character, and mode structure of the transverse susceptibility.
Beyond the spectroscopic relevance, these results may also inform spintronic applications, where tunable magnon dispersions and flat modes could be used for controlled spin transport, information processing, and low-dissipation magnonic devices~\cite{chumak2015magnon, rana2019towards, vzelezny2018spin}.
More broadly, the results provide a concrete framework for relating microscopic magnetic states to experimentally measurable spin dynamics in correlated quantum materials.

\acknowledgments

We thank Claudio Verdozzi and Thorbj{\o}rn Skovhus for useful discussions. 
This work was partially supported by the Wallenberg Initiative Materials Science for Sustainability (WISE) funded by the Knut and Alice Wallenberg Foundation.
EvL acknowledges support from the Swedish Research Council (Vetenskapsrådet, VR) under grant 2022-03090, from the Royal Physiographic Society in Lund and by eSSENCE, a strategic research area for e-Science, grant number eSSENCE@LU 9:1.
HURS acknowledges financial support from the Swedish Research Council (Vetenskapsrådet, VR) grant number 2024-04652 and funding from the European Research Council (ERC) under the European Union’s Horizon 2020 research and innovation programme (grant agreement No.\ 854843-FASTCORR). Some of the computations were enabled by resources provided by the National Academic Infrastructure for Supercomputing in Sweden (NAISS), partially funded by the Swedish Research Council through grant agreement no. 2022-06725. The computations were enabled by resources provided by LUNARC, The Centre for Scientific and Technical Computing at Lund University, under projects LU 2026/7-51 and LU 2026/17-24.
T.F.Y. and B.T.Z. thank the support by NSFC-Young Scientists Fund (No. 12504194), Guangdong Provincial Quantum Science Strategic Initiative (Grant No. GDZX2501004) and Start-up Fund of HKUST(GZ) through grant No. G0104000263.

\bibliography{reference}

\appendix

\section{Details of Interacting Lieb-kagome Phase Diagram}\label{app:phasediagram}

In this section, we show the detailed discussions for the magnetic phase diagram in the Lieb-kagome Hubbard model.
To summarize the key of the HF results: we find that the order parameter $\langle S^z\rangle$ drops quickly near the ferrimagnetic–paramagnetic boundary at small $U$, while at $U=8$ it remains large up to $t'/t\approx0.7$; the off-diagonal block ratio $R_{\rm block}$ (defined by Eq.~\eqref{offratio}) is negligible deep in the ferrimagnet but becomes finite close to the AFM region, indicating spontaneous $\langle c^\dagger_{i\uparrow}c_{j\downarrow}\rangle$ and transverse spin coherence.  Correspondingly, HF produces spin-dependent on-site potentials and band renormalizations that split majority/minority bands, produce band degeneracy in the paramagnet, and give states with mixed $\langle S^z\rangle$ and sizable $\langle S^{x,y}\rangle$ weight (canted phases) at large $U$ and $t'$.

\subsection{Ferrimagnetic Magnetization and Charge Redistribution}
\label{sec:results:orderparameter}

\begin{figure}
\includegraphics[width=0.45\textwidth]{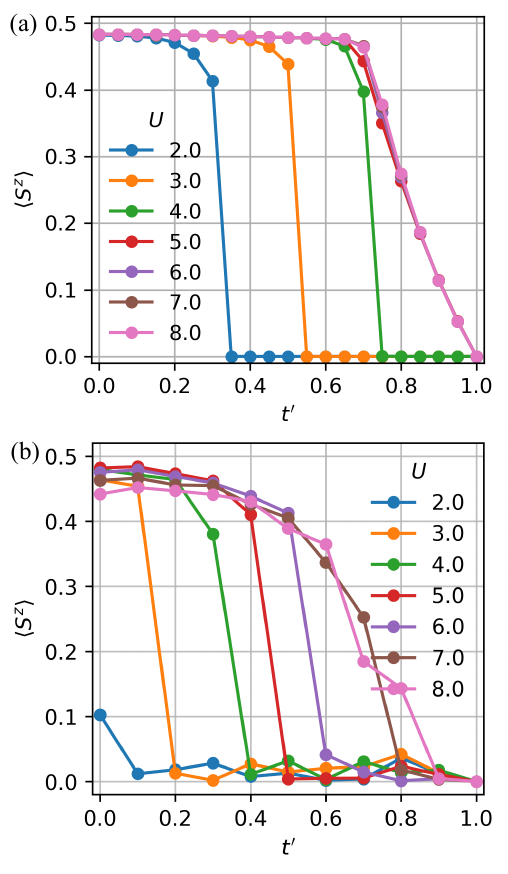}
\caption{Average magnetization per unit cell $\langle S^z\rangle$ as a function of $t^\prime$ with different values of $U$ computed using (a) the HF approximation and (b) the dynamical mean-field theory (DMFT).}
\label{fig:magnetization_t}
\end{figure}

\begin{figure}
\includegraphics[width=0.45\textwidth]{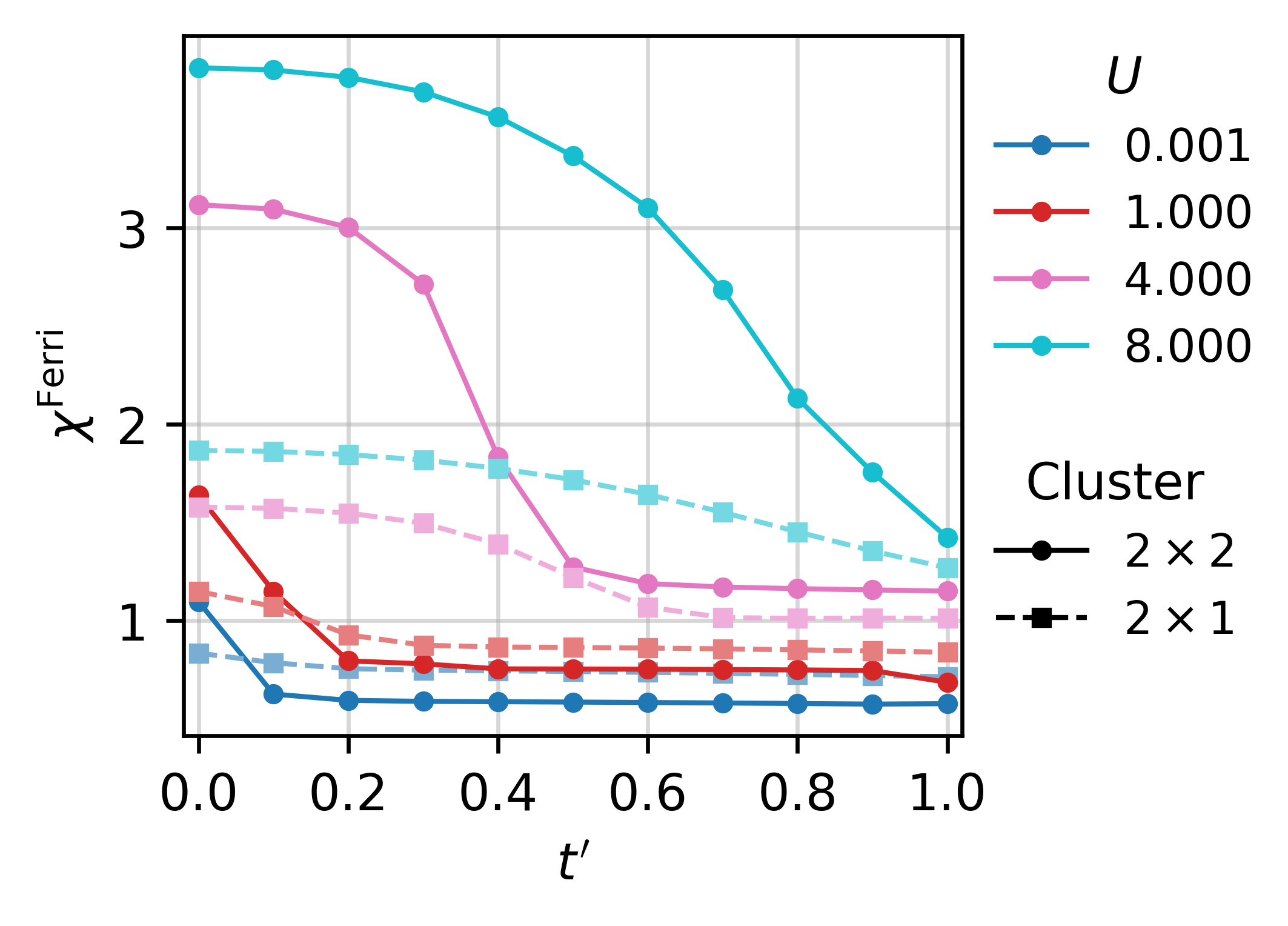}
\caption{The ED ferrimagnetic correlation function $\chi^\text{Ferri}=\sum_{ij} (-1)^i(-1)^j\langle S^z_i S^z_j \rangle$, where $(-1)^j=+1$ for A sites and -1 for B or C sites. The data is computed from the $2\times2$ and $2\times1$ clusters.}
\label{fig:ED_magnetization_t}
\end{figure}

\begin{figure}
\includegraphics[width=0.45\textwidth]{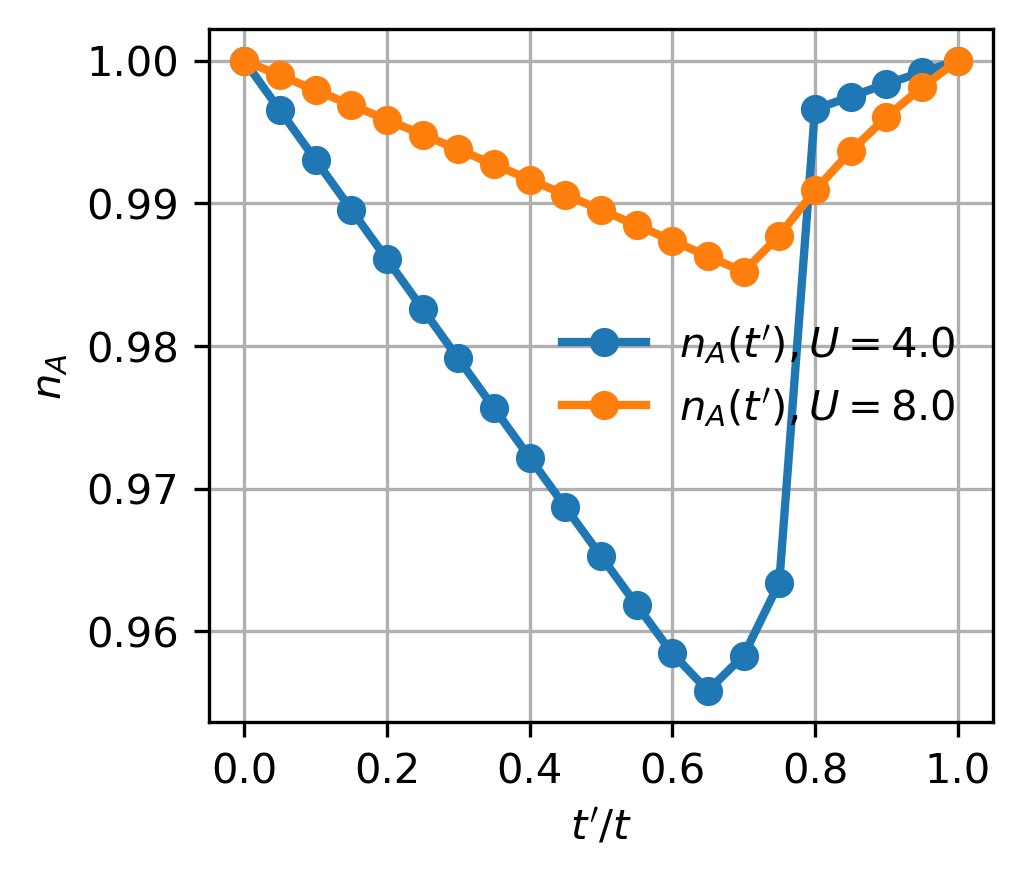}
\caption{Sublattice occupation at site A $n_A$ as a function of $t'$. Results are shown for interaction strengths $U=4.0$ (blue) and $U=8.0$ (orange) at half-filling in the HF approximation.}
\label{fig:na_t}
\end{figure}

\begin{figure}
\includegraphics[width=0.45\textwidth]{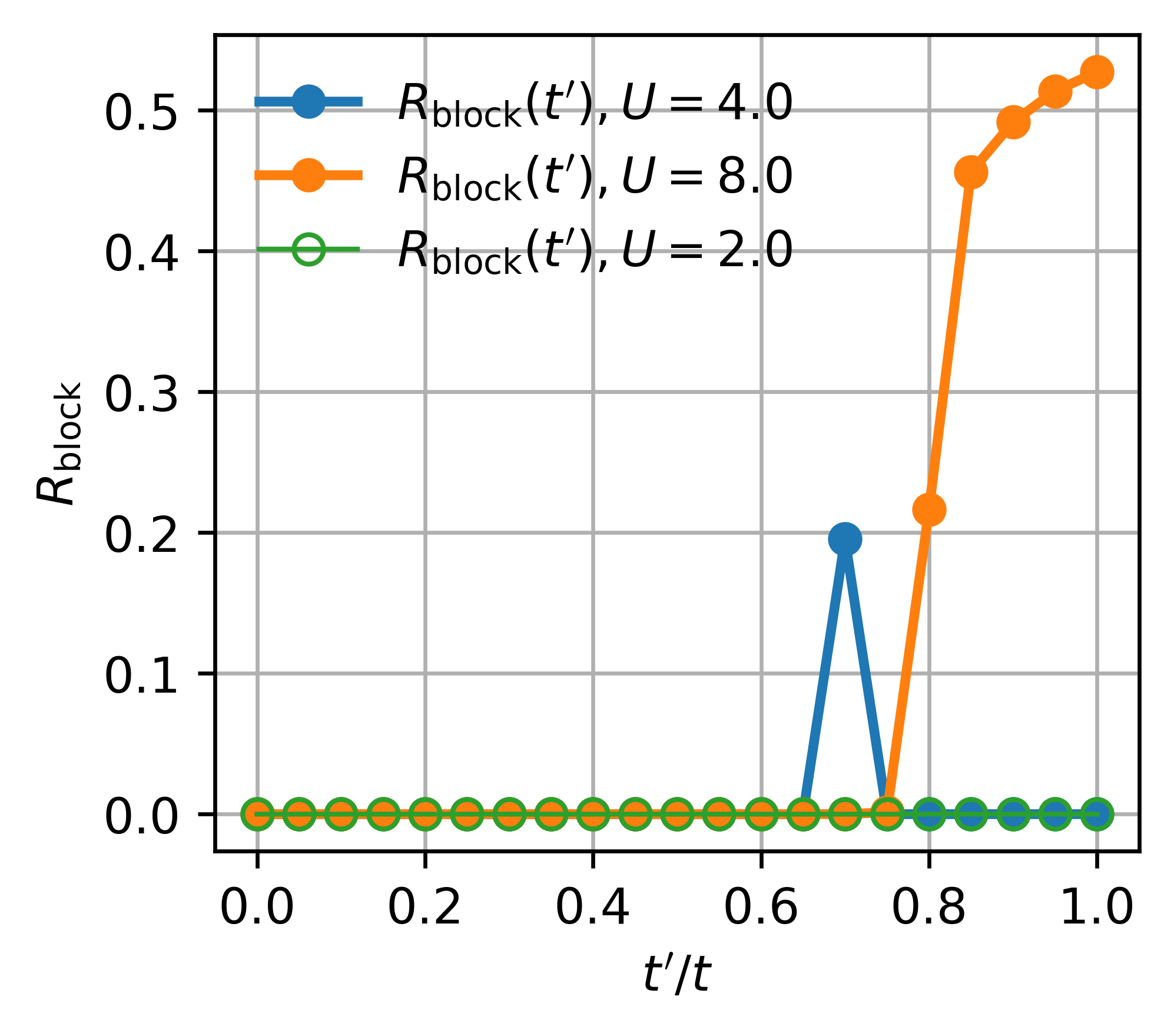}
\caption{Ratio of off-diagonal block to diagonal block magnitudes of the density matrix ($R_\mathrm{block}$, defined by Eq.~\ref{offratio}) as a function of $t'$ in the HF approximation.}
\label{fig:ratio}
\end{figure}

To get a deeper understanding of the phase transitions, we can study the site- and spin-dependent occupation numbers $\langle n_{i\sigma}\rangle$, since they play a central role in the HF approximation for the Hubbard model. 

We start with the average magnetization per unit cell $\langle S^z\rangle=\frac{1}{2}\sum_i \langle n_{i\uparrow}-n_{i\downarrow}\rangle $ in the HF framework. This magnetization is the order parameter of the ferrimagnetic phase and is shown as a function of $t^\prime$ in Fig.~\ref{fig:magnetization_t}(a). It is also shown in color in Fig.~\ref{fig:phaset}(a). For small values of $U$, the magnetization drops quickly close to the ferrimagnetic-paramagnetic phase boundary. At larger values of $U$, there is a more gradual decrease for $t'/t\geq 0.7$, with the magnitude of the ferrimagnetic order parameter going continuously to zero as the antiferromagnetic phase is approached. 
As a contrast, Fig.~\ref{fig:magnetization_t}(b) shows the corresponding DMFT results for the same model. While the overall trend remains qualitatively consistent with the HF phase diagram, two notable differences emerge: (i) the critical interaction strength required to stabilize magnetization is larger for every value of $t'$, and (ii) the effect of geometric frustration occurs at smaller $t'$. Both effects are because HF tends to overestimate the stability of magnetic ordered states.

In ED, there is no spontaneous symmetry breaking and $\langle S^z\rangle=0$ holds at all temperatures and interaction strengths. To probe the strength of ferrimagnetic fluctuations, it is useful to look at the correlation function $\chi^\text{Ferri}$ shown in Fig.~\ref{fig:ED_magnetization_t}, which is for ferrimagnetism what the staggered correlation function is for antiferromagnetism. If one could simulate large systems with ED, the phase diagram could be extracted from the scaling of this correlation function: In a ferrimagnetic phase, $\chi^\text{Ferri}\sim L^d$ where $L$ is the linear size and $d$ is the dimension. $\chi^\text{Ferri}$ goes to a constant independent of $L$ in a nonmagnetic phase, with saturation occurring when the correlation length $\xi \ll L$. The Mermin-Wagner theorem in 2D would show up as a saturation of $\chi^\text{Ferri}$ at very large values of $L$.

In addition to the magnetization per unit cell, another order parameter is the occupation of the corner site, $n_A=\langle n_{A,\uparrow} +n_{A,\downarrow}\rangle$, shown in Fig.~\ref{fig:na_t} for Hartree-Fock. Since the total filling is three by construction, $n_B=n_C=(3-n_A)/2$. In the kagome limit at $t'/t=1$, all sites are equivalent so $n_A=1$. In the Lieb limit, $t'=0$, particle-hole symmetry ensures that $n_A=1$. In between, the HF results show that $n_A$ dips slightly below 1, showing that it is beneficial to have a non-uniform charge distribution. For larger $U$, this effect is smaller, since the Hubbard interaction disfavors charge fluctuations.

Besides the densities, i.e., the diagonal elements of the density matrix, the off-diagonal elements of the HF density matrix are also important. In particular, at large interactions, we find a solution with finite $\langle c^\dagger_{i\uparrow} c_{j\downarrow}\rangle$, corresponding to a spontaneously emerging spin-orbit coupling. In that case, $\av{S^z}$ is no longer a good quantum number and we have complete breaking of the spin symmetry.
To quantify the weight of $\langle c^\dagger_{i\uparrow} c_{j\downarrow}\rangle$ within the HF solutions, we define the off-diagonal-block ratio $R_\mathrm{block}$ as 
\begin{equation}\label{offratio}
R_\mathrm{block}=\frac{\sum_{i}\sum_{j}|\rho_{i\uparrow, j\downarrow}|+\sum_{i}\sum_{j}|\rho_{i\downarrow,j\uparrow}|}{\sum_{i}\sum_{j}|\rho_{i\uparrow,j\uparrow}|+\sum_{i}\sum_{j}|\rho_{i\downarrow,j\downarrow}|},
\end{equation}
which is determined by the absolute values of the HF density matrix elements. The results of $R_\mathrm{block}$ as a function of $t'$ with $U=4.0$ and $U=8.0$ are shown in Fig.~\ref{fig:ratio}.
\begin{figure}
\includegraphics[width=0.45\textwidth]{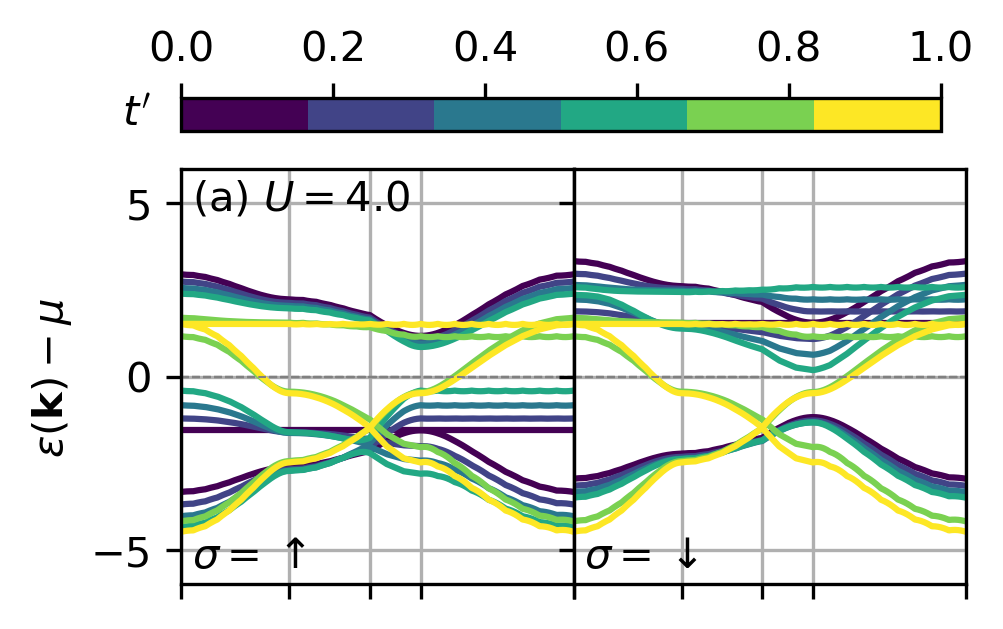}\\
\includegraphics[width=0.45\textwidth]{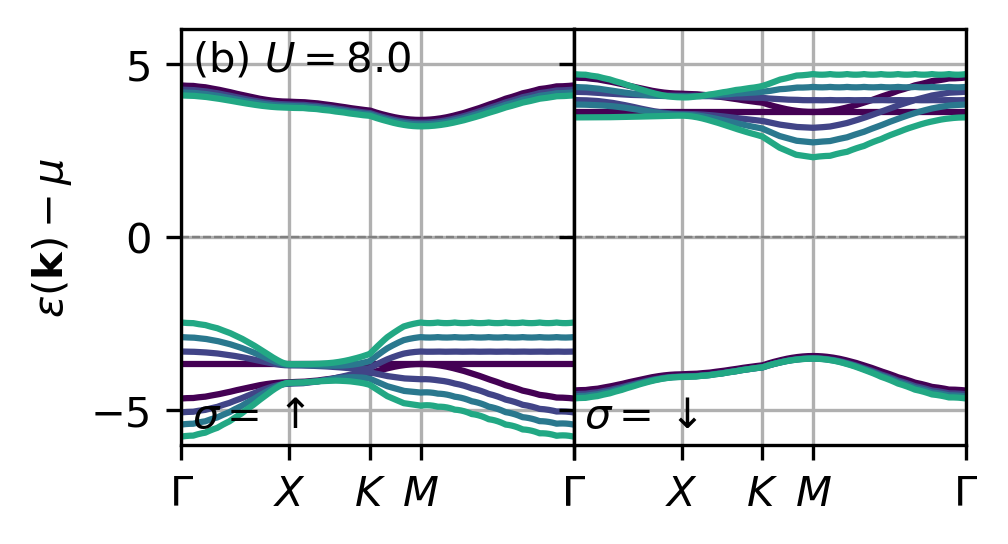}
\caption{Evolution of band structure with $t'$ at (a) $U=4.0$, $0.0\leq t'\leq 1.0$ and (b) $U=8.0$, $0.0\leq t'\leq 0.6$ along the high-symmetry path.}
\label{fig:bandstructure}
\end{figure}
From Fig.~\ref{fig:ratio}, we see that the off-diagonal components of the HF density matrix are zero in the ferrimagnetic phase for both $U=4.0$ and $8.0$. However, by increasing $t'$ at $U=4.0$, the ratio $R_\mathrm{block}$ becomes finite at a point close to the AFM state and collapses to zero in the paramagnetic state, indicating a complete loss of in-plane spin-polarization in the paramagnetic phase. 
For $U=8.0$, $R_\mathrm{block}$ increases quickly between the ferrimagnetism-antiferromagnetism phase transition, indicating the strong in-plane polarization in this state. This will also be examined in Fig.~\ref{fig:SxSyweightU8t1}. 

We note that a purely Hartree approximation does not produce antiferromagnetic symmetry breaking at large~$U$. Instead, it drives the kagome lattice toward a ferrimagnetic state with strong $S^{z}$ polarization, reflecting both the odd number of bands and the inability of the Hartree term to capture geometric frustration. In contrast, the inclusion of the Fock terms, which is activated by a small transverse seed field giving the finite values of $R_{\mathrm{block}}$, allows transverse spin coherences that lower the energy and stabilize the antiferromagnetic configuration observed in our calculations.

\subsection{Band Structure Renormalization}
\label{sec:results:bandstructure}

In the HF approximation, the ferrimagnetism and charge redistribution are felt by the electrons in the form of on-site potentials and renormalization of the hopping, both spin-dependent, which lead to a renormalization of the band structure. The evolution of the band structure with $t'$ is shown in Fig.~\ref{fig:bandstructure}, where we have restricted the plot to the regime where $S^z$ remains a good quantum number. 

At moderate interaction $U=4.0$, starting from $t'=0$, the spin–up and spin–down bands are separated by the orbital-dependent exchange splitting. The flat band remains flat in the presence of the interaction, but the majority flat band is below the Fermi level, while the minority flat band is above the Fermi level. 
As $t'$ increases, the phase transition to paramagnetism happens between $t'=0.6$ and $t'=0.8$, as shown in Figs.~\ref{fig:phaset} and \ref{fig:magnetization_t}. As a result, the spin-up and spin-down bands become degenerate in the paramagnetic phase.

In the strong-coupling case with $U=8.0$, the band gap in the ferrimagnetic case is larger because of the stronger Hubbard interaction. As a result, the ferrimagnet is also more robust, and the order parameter $\langle S^z\rangle$ remains close to its maximum value ($\langle S^z \rangle = 1/2$) up to
$t'/t \lesssim 0.7$.

\begin{figure}
\includegraphics[width=0.45\textwidth]{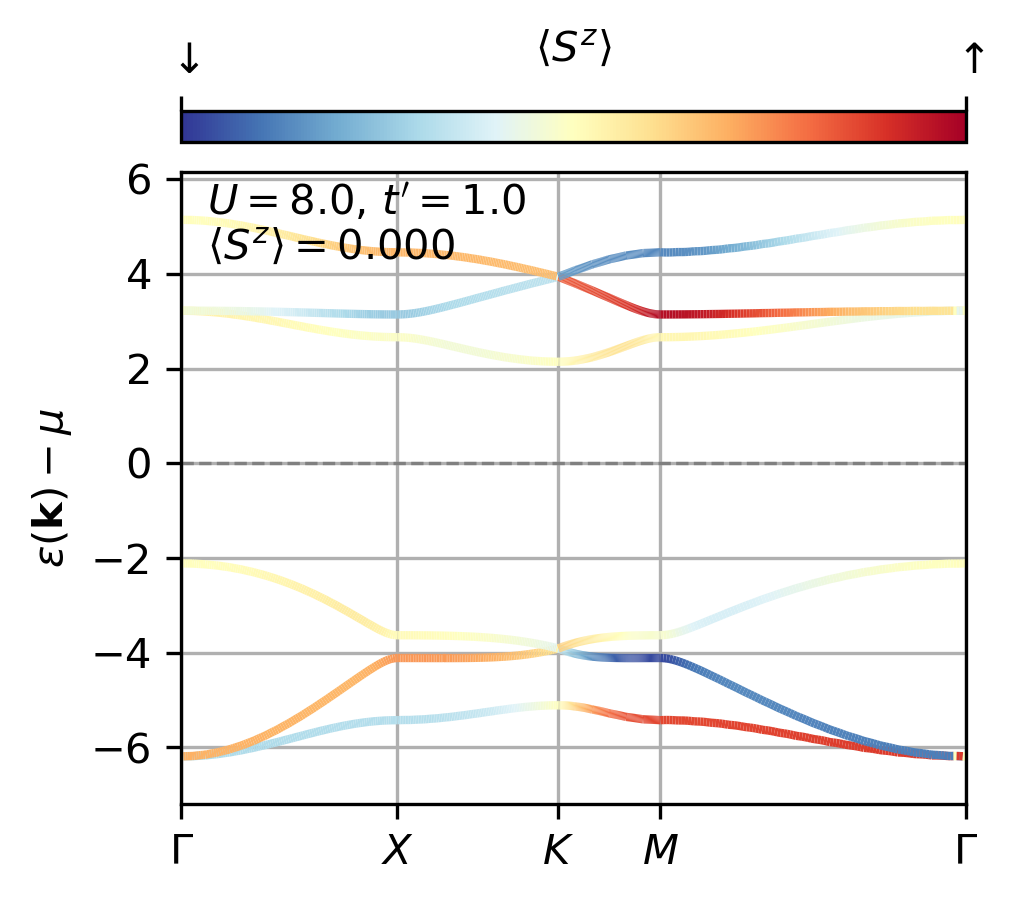}
\caption{Band structure colored by $\langle S^z\rangle$ of each eigenstate with $U=8.0$ and $t'=1.0$. The color scale represents the spin-up weight $\sum^3_{i=1}|\psi_i|^2$, ranging from fully spin-down (blue) to fully spin-up (red).}
\label{fig:spinweightU8t1}
\end{figure}

\begin{figure}
\includegraphics[width=0.45\textwidth]{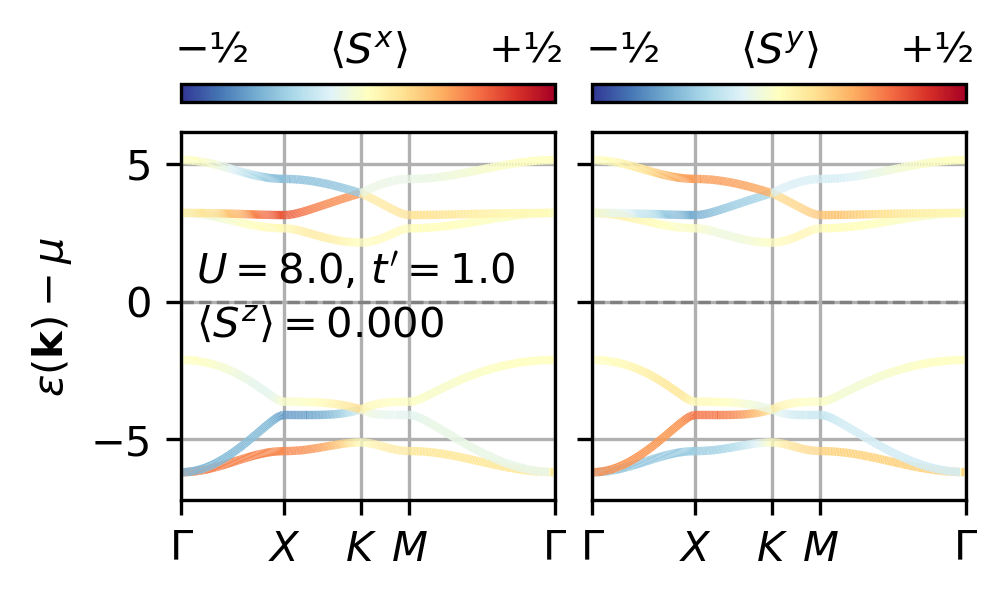}
\caption{Band structure colored by the transverse spin‐polarization components. Left: $\langle S^x\rangle=\sum_i\langle\psi_i|S^x|\psi_i\rangle$ ranged from fully $-\frac{1}{2}$ (blue) to $\frac{1}{2}$ (red). Right: $\langle S^y\rangle=\sum_i\langle\psi_i|S^y|\psi_i\rangle$ on the same color scale. For both panels, we set $U=8.0$ and $t'=1.0$.}
\label{fig:SxSyweightU8t1}
\end{figure}

For larger $t'$, $S^z$ is no longer a good quantum number, and the bands have partial $S^z$ weights as shown in Fig.~\ref{fig:spinweightU8t1}. Simultaneously, the bands become polarized in the $S^x$ and $S^y$ directions, as shown in Fig.~\ref{fig:SxSyweightU8t1}. In other words, the spin polarization vector of the bands rotates along the Bloch sphere, indicating the phase transition from canted ferrimagnetism, which is also observed in the quarter-filled Lieb lattice~\cite{nikolaenko2025canted}, to the canted antiferromagnetism. In this sense, the reduction of $\langle S^z\rangle$ with $t'$, visible in Fig.~\ref{fig:magnetization_t}, happens because the spin polarization of the filled bands changes continuously with $t'$. This also means that the concepts of minority and majority bands break down.

\subsection{Longitudinal Susceptibility}

\begin{figure}
    \centering
    \includegraphics[]{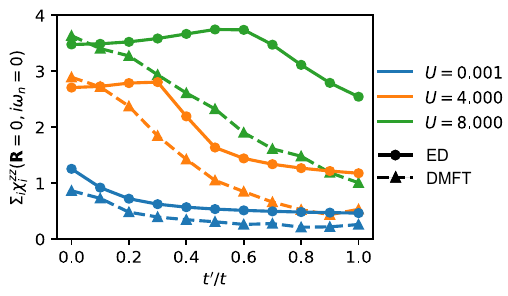}
    \caption{The static local longitudinal susceptibility $\chi^{zz}(\mathbf{R}=0,i\omega_n=0)$ as functions of $t'$ with different values of $U$ at $\beta=10$, computed by exact diagonalization (solid lines with rounded markers) and dynamical mean-field theory (dashed lines with triangular markers).}
    \label{fig:chi_SzSz_ED_DMFT}
\end{figure}

Besides the transverse susceptibility $\chi^{+-}$, which measures the transverse spin-flip dynamics and is directly related to the magnon and spin-excitation spectrum, we also present the longitudinal susceptibility $\chi^{zz}$ in this section. Different from $\chi^{+-}$, $\chi^{zz}$ measures fluctuations in the magnitude of the local magnetization, capturing the dominant magnetic response and indicating the corresponding magnetic order.

Figure~\ref{fig:chi_SzSz_ED_DMFT} compares the static local longitudinal susceptibility $\chi^{zz}(\mathbf{R}=0,i\omega_n=0)$ obtained from ED and DMFT for several interaction strengths and values of $t'$. In all cases, increasing $U$ strongly enhances the low-frequency longitudinal response, reflecting the growth of local moment fluctuations in the correlated regime. The dependence on $t'$ is more method dependent: in DMFT, $\chi^{zz}(0)$ decreases monotonically as $t'$ increases for all three interaction strengths, indicating that larger $t'$ generally weakens the local longitudinal response. In ED, the trend is nonmonotonic at intermediate and strong coupling. For $U=4$ and $U=8$, $\chi^{zz}(\mathbf{R}=0,i\omega_n=0)$ remains large over a broad range of small and intermediate $t'$ and only drops appreciably as the system approaches the kagome-lattice limit, where geometric frustration suppresses the response.

Comparing ED and DMFT at the smallest $U$, where both methods are essentially exact, it is important to note that DMFT still simulates a large lattice and is therefore not identical to the ED results for the small cluster. For larger $U$, the deviations between ED and DMFT are due to both finite-size effects in ED and the approximate nature of DMFT. Overall, the longitudinal susceptibility shows enhanced magnetic fluctuations in the same regimes as the Hartree-Fock and DMFT phase diagrams. 

\subsection{Chemical Potential}

\begin{figure}[ht!]
\includegraphics[width=0.45\textwidth]{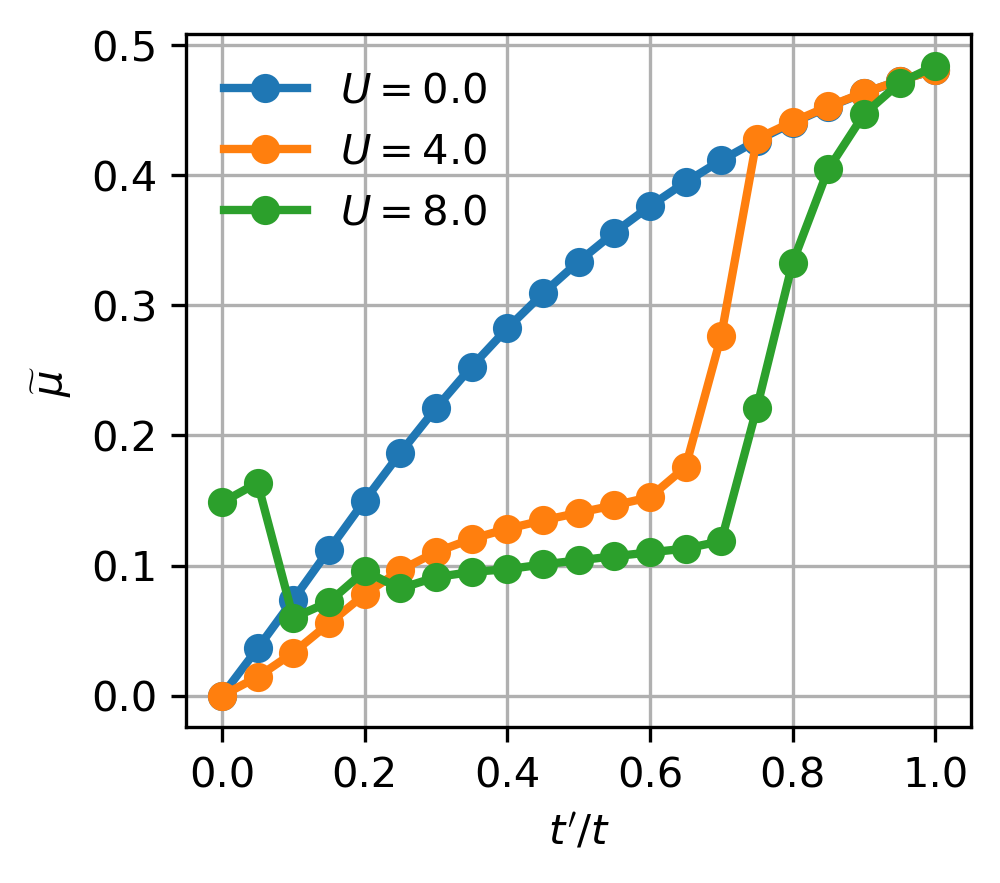}
\caption{The reduced chemical potential, $\widetilde\mu=\mu-\frac{1}{2}U$, that leads to half-filling, plotted as a function of $t'$ for $U=0.0$, $4.0$, and $8.0$.}
\label{fig:chemical_mu_t}
\end{figure}

Figure~\ref{fig:chemical_mu_t} shows how the reduced chemical potential $\widetilde\mu=\mu-\frac{1}{2}U$ in the Hubbard model that leads to half-filling evolves in the Lieb-kagome lattice by tuning $t'$, for $U=0$, $4$, and $8$.
In the non-interacting limit, $\widetilde\mu$ increases smoothly with $t'$ and reaches approximately 0.48 near the kagome limit. For moderate and strong interactions, the shift in chemical potential similarly approaches $\sim0.48$ as $t'\rightarrow1$, which gives a similar result by DQMC at the temperature $\beta=10$~\cite{lima2023magnetism}.

For $U=4$ and $U=8$, the dependence of $\mu$ on $t'$ is no longer smooth, instead showing jumps and sudden changes in slope. These changes occur when a qualitatively different mean-field solution is found. For $U=4$, the jump occurs around $t'=0.75$ where the ferrimagnetic-paramagnetic transition takes place. For $U=8$, we see several changes even within the ferrimagnetic phase. The discontinuities in the chemical potential indicate metal-insulator transitions~\cite{ono2001critical}.

This HF behavior contrasts with the smoother trend observed in DQMC. 
This difference reflects the tendency of HF to overestimate order and produce first‑order jumps, whereas the exact solution does not have phase transitions in 2d, instead showing smooth cross-overs between nearly ordered phases. Still, the HF is useful to signal what kind of fluctuation is dominant.

\section{Non-interacting Model}
\label{app:noninteracting}

\begin{figure}
\includegraphics[width=0.45\textwidth]{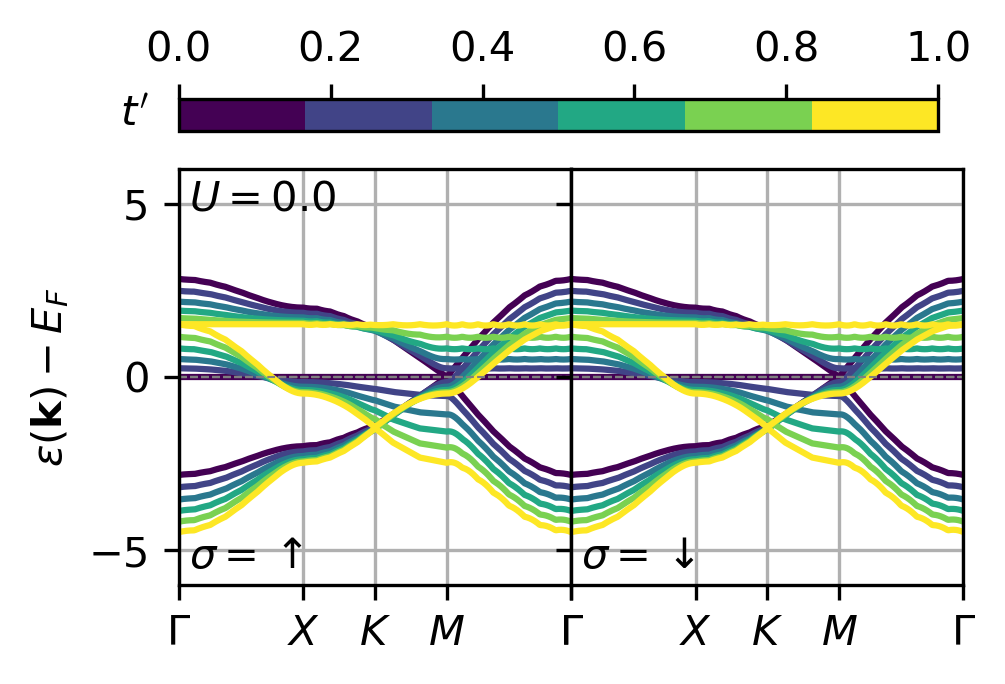}
\caption{Evolution of band structure as a function of $t'$ at $U=0.0$}
\label{fig:bandstructureU0}
\end{figure}

\begin{figure}
\includegraphics[width=0.235\textwidth]{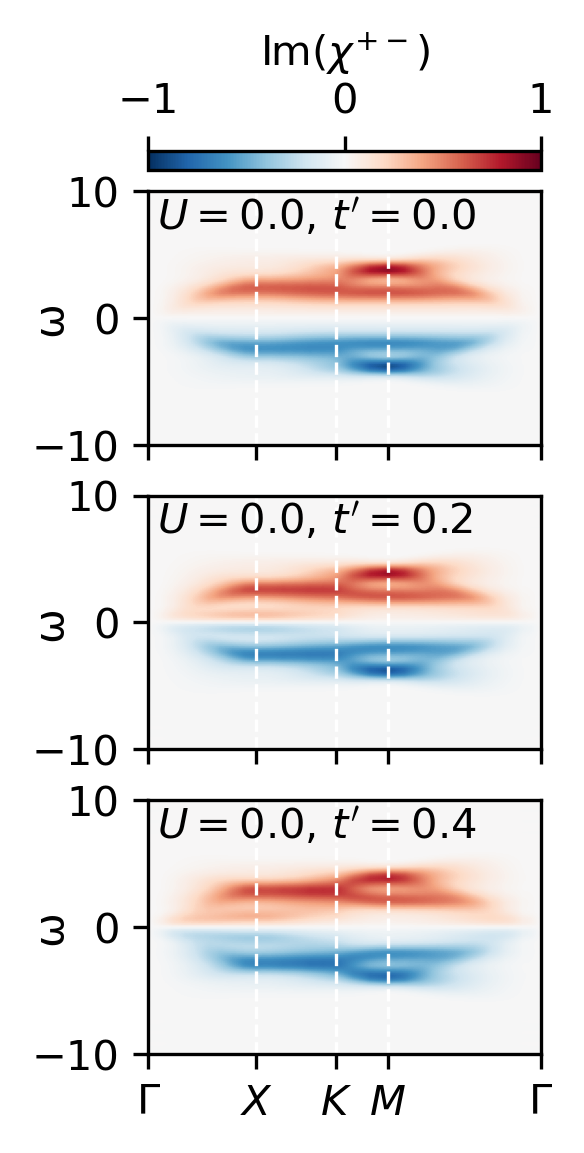}
\includegraphics[width=0.235\textwidth]{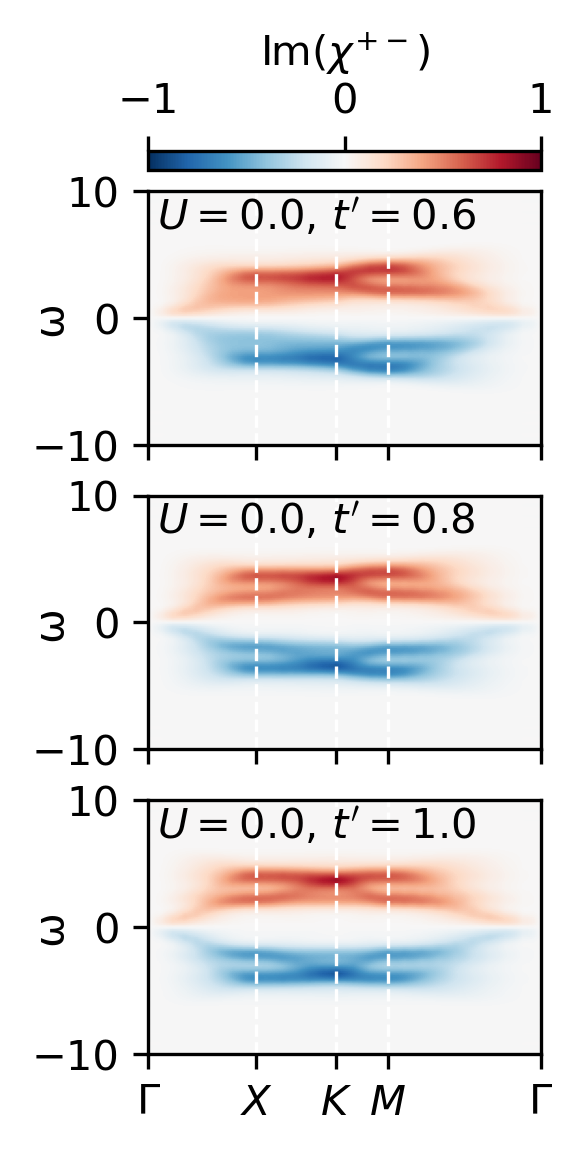}
\caption{Magnetic susceptibility $\Im\chi^{+-}(\omega,\qv)$ of the non-interacting system.}
\label{fig:susceptibility0.0}
\end{figure}

\begin{figure}
\includegraphics[width=0.5\textwidth]{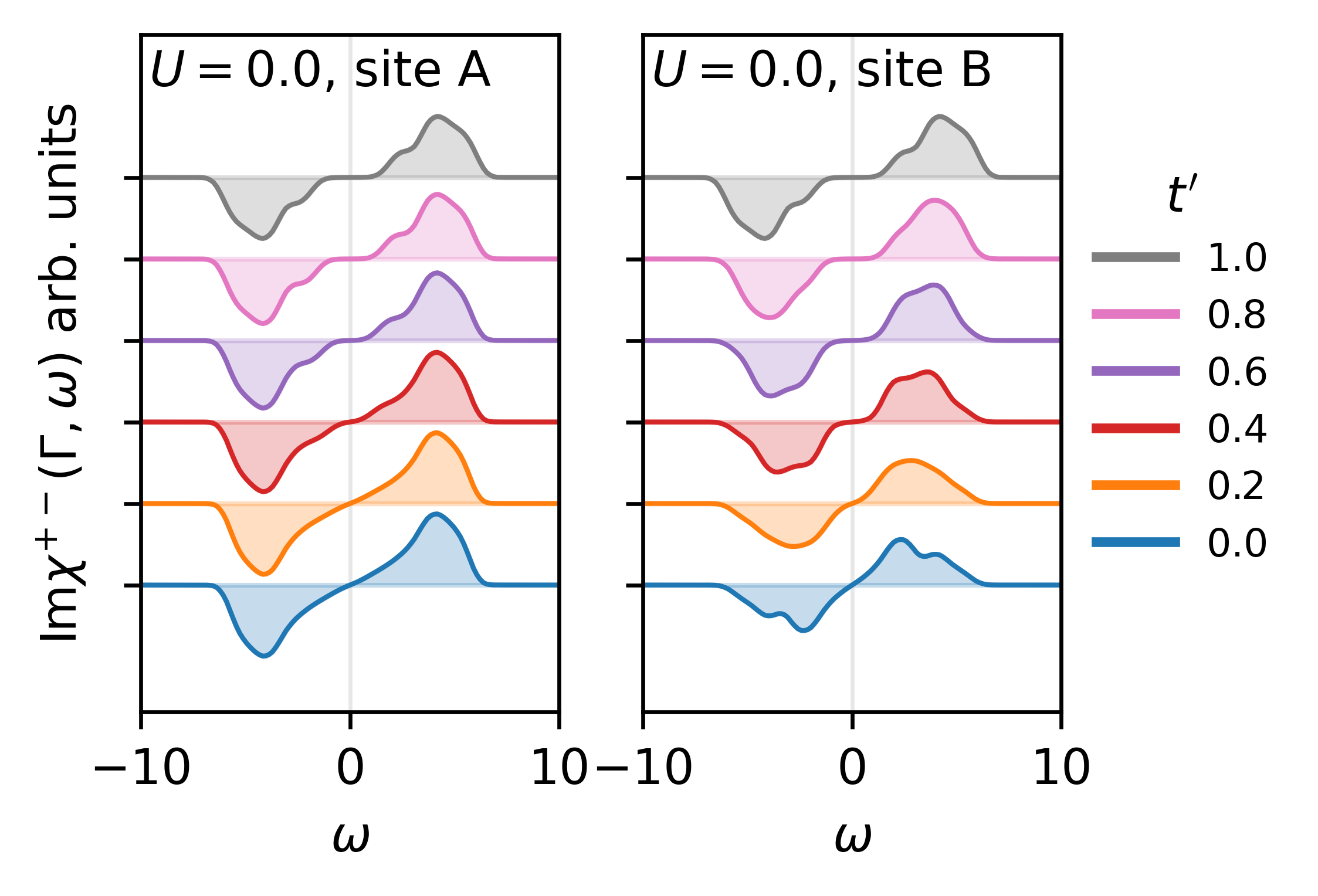}
\caption{Cross-section of the magnetic susceptibility at $\Gamma$ for interaction strength $U=0.0$, shown separately for site A (left) and site B (right). }
\label{fig:susceptibility:gammaU0}
\end{figure}

The non-interacting system is paramagnetic and has a band structure which evolves smoothly with $t'$, as shown in Fig.~\ref{fig:bandstructureU0}. Note that there is a perfectly flat band for $t'=0$ and $t'=t$ only. The corresponding (bare) susceptibility is shown in Fig.~\ref{fig:susceptibility0.0}. Due to the lack of $SU(2)$ symmetry breaking, there is a paramagnon Goldstone mode near the $\Gamma$ point and there is a perfect symmetry between positive and negative energies. As $t'/t\to 1$, the $X-K$ and $X-M$ sections become equivalent, while they are clearly distinct for $t'=0$.

In the paramagnetic Lieb lattice, the flat band is fixed at the Fermi level, and the other two bands are particle-hole symmetric, so there are relatively few different transitions, and the susceptibility is dominated by a single mode for most of the Brillouin zone. The reduced symmetry for $t'>0$ leads to additional modes appearing in $\chi$. In particular, at finite $t'$, the previously flat band now crosses the Fermi level, and transitions within this almost flat band give rise to low-energy features in the susceptibility.

Moving on to the site-resolved $\Gamma$-point susceptibility, Fig.~\ref{fig:susceptibility:gammaU0} shows that A site, which is the minority site in the Lieb lattice, is relatively insensitive to $t'$, while the B site is more sensitive.

\section{Phase Diagram at Lower Temperature}
\label{app:temperature}

\begin{figure}
\includegraphics{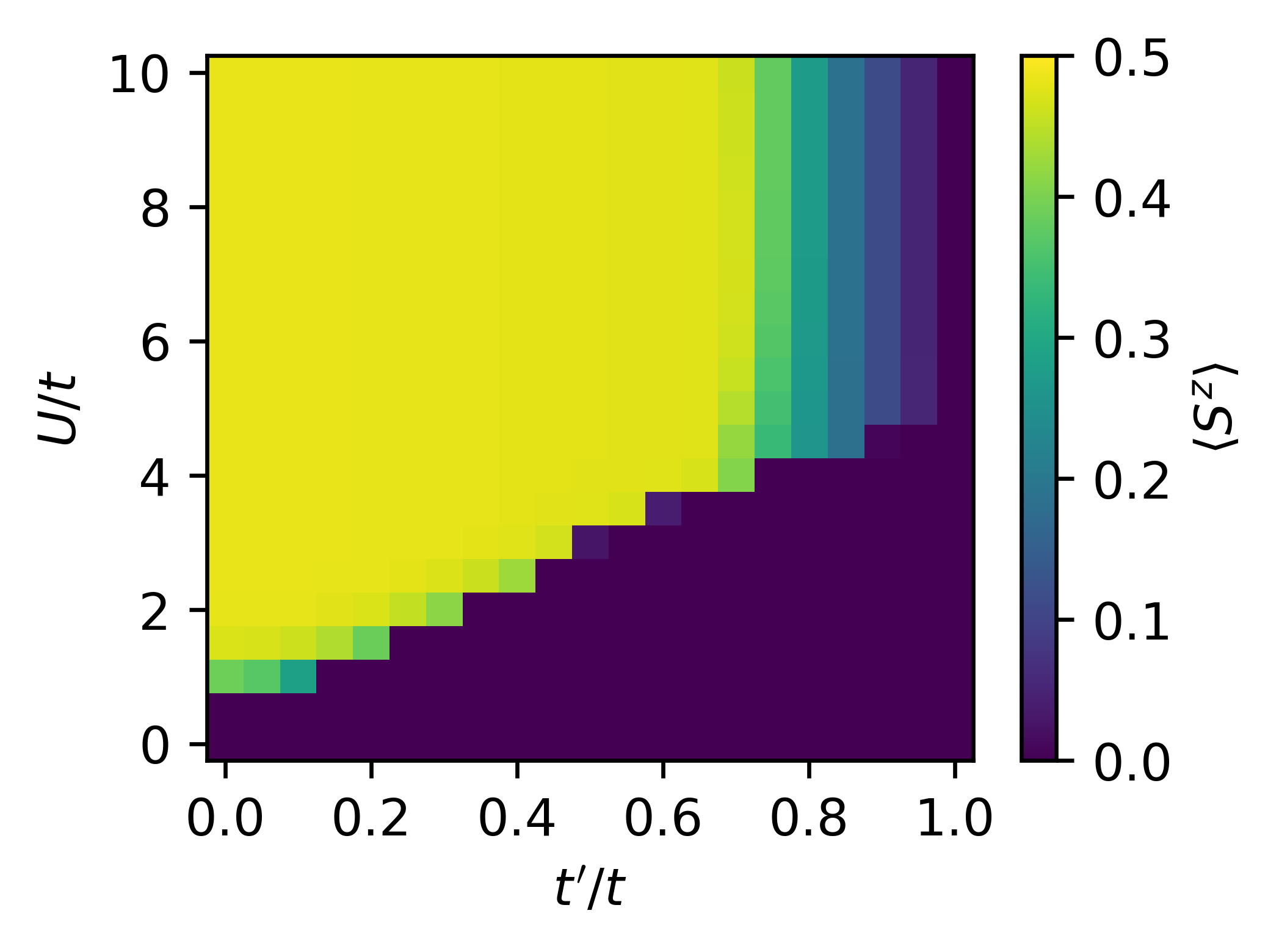}
\caption{Phase diagram in the HF approximation of the Lieb-kagome lattice at the temperature $\beta=50$. The heatmap presents the magnetization as a function of $U$ and $t^\prime$.}
\label{fig:phaset50}
\end{figure}

Figure~\ref{fig:phaset50} presents the same figure as Fig.~\ref{fig:phaset} but with a lower temperature $\beta=50$ for comparison. We can observe that the basic feature of the magnetic phase transition is identical for the temperature we mainly investigate in the main text, compared to the lower temperature, which demonstrates that we already approach the low-temperature limit even at $\beta=10$.

\begin{figure}
    \centering
    \includegraphics{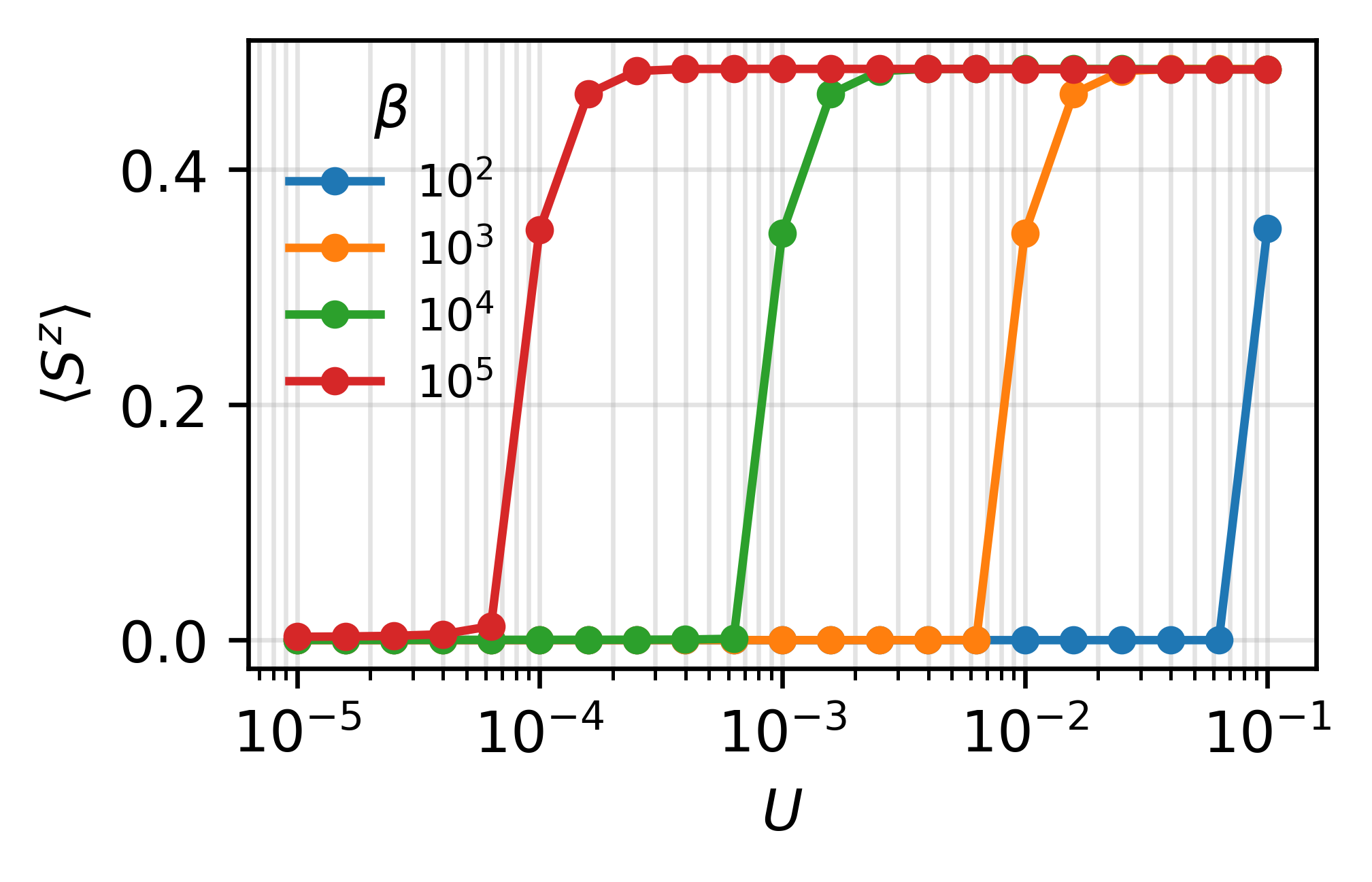}
    \caption{Magnetization $\av{S^z}$ as functions of $U$ at different inversed temperatures $\beta$ in the Lieb lattice limit.}
    \label{fig:HF_m_U_beta}
\end{figure}

The scaling of $U_P(T)$ in the Lieb lattice limit ($t'=0$), can be understood from the generalization of the Stoner criterion to finite temperature, for a flat band. For the flat band in the Lieb lattice, the energy associated with an occupation $\delta n$ of the flat band is $\frac{1}{2} U (\delta n)^2$ since this wave function is spread out over two sites, so the interaction that enters the Stoner criterion is $\frac{U_P}{2}$. The density of states at the Fermi level, $A(E_F)$, in the Stoner criterion is replaced by $-\int A(E) \frac{df_\text{FD}}{dE} dE$ at finite temperature, which gives
\begin{align}
    1 &= -\frac{U_P}{2} \int A(E) \frac{df_\text{FD}}{dE} dE=  \frac{\beta U_P}{2} \int A(E) f_\text{FD} (1-f_\text{FD}) dE \notag\\
    1&\overset{\text{Lieb}}{=} \frac{\beta U_P}{8},
\end{align}
so that $U_P \sim 1/\beta$, consistent with the numerical results in Fig.~\ref{fig:HF_m_U_beta} and with $U_P \to 0$ in the zero-temperature limit.

\section{DMFT Calculations}
\label{sec:DMFTmethod}
We also solved the Hubbard Hamiltonian using standard dynamical mean-field theory (DMFT) at the single-particle Green's function level, implemented in the Toolbox for Research on Interacting Quantum Systems (TRIQS)~\cite{parcollet2015triqs} and its Two-Particle Response Function toolbox (TPRF)~\cite{tprf}. 
The DMFT self-energy was calculated self-consistently using the continuous-time hybridization expansion quantum Monte Carlo (CT-HYB) solver~\cite{cthyb} with a spin-diagonal Green's function. 
The local physical observables, including the density matrix $\rho_{i\sigma,j\sigma'}$ and lattice magnetization $\av{S^z_i}$, are evaluated from the single-particle Green's function solved by the lattice Dyson equation in DMFT. The dynamic spin susceptibility in the $S^zS^z$ channel can be measured directly in the impurity solver.
We perform the calculation within the parameter range of $0<U\leq10$ and $0\leq t'<1$ at $\beta=10$ and half-filling, and assume that $\av{S^z}=0$ for $U=0$ as the non-interacting limit and $\av{S^z}=0$ for $t'=1$ (kagome lattice) as the fully frustrated configuration.

\section{ED calculations}
\label{sec:EDmethod}
For the exact diagonalization, we use a cluster with 12 sites in total, with $2\times 2$ unit cells of the Lieb-kagome lattice. Periodic boundary conditions are used. The exact diagonalization is performed using EDLib~\cite{iskakov2018exact}. This code produces both the low-lying eigenvalues and eigenvectors of the cluster as well as thermal expectation values in the grand canonical ensemble. For the grand canonical ensemble, we use $\mu=U/2$, which gives a filling of approximately 1 electron per site for the clusters considered here (this choice gives filling exactly 1 at $t'=0$). The exact solution of a finite cluster will not show spontaneous symmetry breaking, so it is not possible to produce a phase diagram. However, the ferrimagnetic correlation function $\chi^\text{Ferri}=\sum_{ij} (-1)^i(-1)^j\langle S^z_i S^z_j \rangle$, where $(-1)^j=+1$ for A sites and -1 for B or C sites, gives an indication of the strength of ferrimagnetic correlations of the finite cluster. If one could perform finite-size scaling, this correlation function would grow linearly with system size in a ferrimagnetically ordered phase while plateauing to a finite value otherwise.

\end{document}